\documentclass[aps,prc,twocolumn,superscriptaddress,showpacs,showkeys,amsmath,amssymb,floatfix]{revtex4}
\usepackage{booktabs,graphicx,mathrsfs,verbatim}
\usepackage[usenames,dvipsnames]{color}
\usepackage{multirow}
\usepackage{ulem}
\usepackage{color}
\usepackage[sort&compress]{natbib} 
\newcommand{\valencia}{\affiliation{Instituto de F{\'i}sica Corpuscular, CSIC-Universidad de Valencia, E-46071 Valencia, Spain}}
\newcommand{\osakadep}{\affiliation{Department of Physics, Osaka University, Toyonaka, Osaka 560-0043, Japan}}
\newcommand{\bordeaux}{\affiliation{Centre d'Etudes Nucl{\'e}aires de Bordeaux Gradignan, CNRS/IN2P3 - Universit{\'e} de Bordeaux, 33175 Gradignan Cedex, France}}
\newcommand{\surrey}{\affiliation{Department of Physics, University of Surrey, Guildford GU2 7XH, Surrey, UK}}
\newcommand{\debrecen}{\affiliation{Inst. of Nuclear Research of the Hung. Acad. of Sciences, Debrecen, H-4026, Hungary}}
\newcommand{\instanbul}{\affiliation{Department of Physics, Istanbul University, Istanbul, 34134, Turkey}}
\newcommand{\caen}{\affiliation{Grand Acc{\'e}l{\'e}rateur National d'Ions Lourds, CEA/DSM - CNRS/IN2P3, BP 55027, F-14076 Caen, Cedex 5, France}}
\newcommand{\osakarcnp}{\affiliation{Research Center for Nuclear Physics, Osaka University, Ibaraki, Osaka 567-0047, Japan}}
\newcommand{\chile}{\affiliation{Comisi{\'o}n Chilena de Energ{\'i}a Nuclear, Casilla 188-D, Santiago, Chile}}
\newcommand{\belgium}{\affiliation{SCK.CEN, Boeretang 200, 2400 Mol, Belgium}}
\newcommand{\argonne}{\affiliation{Physics Division, Argonne National Laboratory, Argonne, Illinois 60439, USA}}

\begin{document} 

\title{Beta decay of the exotic $T_z$ = -2 nuclei $^{48}$Fe, $^{52}$Ni and $^{56}$Zn}

\author{S.~E.~A.~Orrigo}\email{sonja.orrigo@ific.uv.es}\valencia
\author{B.~Rubio}\valencia
\author{Y.~Fujita}\osakadep\osakarcnp
\author{W.~Gelletly}\valencia\surrey
\author{J.~Agramunt}\valencia
\author{A.~Algora}\valencia\debrecen
\author{P.~Ascher}\bordeaux
\author{B.~Bilgier}\instanbul
\author{B.~Blank}\bordeaux
\author{L.~C{\'a}ceres}\caen
\author{R.~B.~Cakirli}\instanbul
\author{E.~Ganio{\u{g}}lu}\instanbul
\author{M.~Gerbaux}\bordeaux
\author{J.~Giovinazzo}\bordeaux
\author{S.~Gr{\'e}vy}\bordeaux
\author{O.~Kamalou}\caen
\author{H.~C.~Kozer}\instanbul
\author{L.~Kucuk}\instanbul
\author{T.~Kurtukian-Nieto}\bordeaux
\author{F.~Molina}\valencia\chile
\author{L.~Popescu}\belgium
\author{A.~M.~Rogers}\argonne
\author{G.~Susoy}\instanbul
\author{C.~Stodel}\caen
\author{T.~Suzuki}\osakarcnp
\author{A.~Tamii}\osakarcnp
\author{J.~C.~Thomas}\caen


\begin{abstract}
The results of a study of the beta decays of three proton-rich nuclei with $T_z=\text{-}2$, namely $^{48}$Fe, $^{52}$Ni and $^{56}$Zn, produced in an experiment carried out at GANIL, are reported. In all three cases we have extracted the half-lives and the total $\beta$-delayed proton emission branching ratios. We have measured the individual $\beta$-delayed protons and $\beta$-delayed $\gamma$ rays and the branching ratios of the corresponding levels. Decay schemes have been determined for the three nuclei, and new energy levels are identified in the daughter nuclei. Competition between $\beta$-delayed protons and $\gamma$ rays is observed in the de-excitation of the $T=2$ Isobaric Analogue States in all three cases. Absolute Fermi and Gamow-Teller transition strengths have been determined. The mass excesses of the nuclei under study have been deduced. In addition, we discuss in detail the data analysis taking as a test case $^{56}$Zn, where the exotic $\beta$-delayed $\gamma$-proton decay has been observed.
\end{abstract}

\pacs{
 23.40.-s, 
 23.50.+z, 
 21.10.-k, 
 27.40.+z, 
}

\keywords{$\beta$ decay, decay by proton emission, $\beta$-delayed $\gamma$-proton decay, $^{48}$Fe, $^{52}$Ni, $^{56}$Zn, Gamow-Teller transitions, isospin mixing, isospin symmetry, mass excess, charge-exchange reactions, proton-rich nuclei}

\maketitle

\section{\label{intro}Introduction}

The investigation of nuclear structure far from the valley of nuclear stability is a topic of the utmost importance in contemporary nuclear physics. The experimental challenge of exploring this $Terra\ Incognita$ demands and has driven the construction of a new generation of facilities for the production and acceleration of radioactive ion beams. Following the steady improvement in the range of beams available and their intensities, more and more proton-rich nuclei can be produced up to the proton drip-line, enabling one to perform detailed decay studies and explore new and exotic decay modes \cite{Blank2008}.

Beta-decay spectroscopy is a powerful tool to investigate the structure of exotic nuclei. Beta decay is a weak interaction process mediated by the well-understood $\tau$ and $\sigma\tau$ operators, responsible for the observed Fermi (F) and Gamow-Teller (GT) transitions, respectively. It provides direct access to the absolute values of the corresponding transition strengths, $B$(F) and $B$(GT).

The GT transitions are characterized by an angular momentum transfer $\Delta L=0$ and spin-isospin flip ($\Delta S=1$ and $\Delta T=1$). For F transitions $\Delta S=0$ and $\Delta T=0$. Due to the simple nature of the operator, the GT transitions are an important tool for the study of nuclear structure \cite{Bohr1969,RevModPhys.64.491,Rubio2009,Fujita2011549}, providing information on the overlap between the wave-functions of the parent ground state and the states populated in the daughter nucleus. Moreover, they play an important role in nuclear astrophysics, especially in stellar evolution, supernovae explosions and nucleosynthesis \cite{RevModPhys.75.819}. Since many heavy proton-rich elements are produced in the $rp$-process passing through proton-rich $fp$-shell nuclei, the study of GT transitions starting from unstable proton-rich nuclei is also of crucial importance. Our knowledge of GT transitions when approaching the proton drip-line is still rather incomplete \cite{RevModPhys.75.819}, however, because the production of such nuclei becomes steadily more challenging.

Normally in proton-rich nuclei proton decay is expected to dominate above the proton separation energy. In this context, proton-rich nuclei with the third component of isospin $T_z=\text{-}2$ are of particular interest because their decay may present peculiarities related to the competition between $\beta$-delayed protons and $\beta$-delayed $\gamma$ rays \cite{PhysRevLett.112.222501,Dossat200718}. Moreover recently a rare and exotic decay mode, the $\beta$-delayed $\gamma$-proton decay, has been observed for the first time in the $fp$-shell in the decay of $^{56}$Zn \cite{PhysRevLett.112.222501}. 

In this paper we present the results of a study of the $\beta$ decay of the $T_z=\text{-}2$ nuclei $^{48}$Fe, $^{52}$Ni and $^{56}$Zn. The analysis of the data is described in detail using the $^{56}$Zn case as an example, expanding on some of the procedures already presented in Ref. \cite{PhysRevLett.112.222501}.

The paper is organized as follows. Section \ref{exp} describes the experiment performed at GANIL. Section \ref{analysis} describes the data analysis procedures. Section \ref{results} presents the experimental results for the $\beta$ decays of $^{48}$Fe, $^{52}$Ni and $^{56}$Zn. The determination of the mass excesses of the nuclei under study is addressed in Section \ref{masses}. Finally, Section \ref{concl} gives our conclusions.

\section{\label{exp}Experimental set-up}

The experiment to study the decay of $^{48}$Fe, $^{52}$Ni and $^{56}$Zn was performed at the LISE3 facility of GANIL (France) \cite{Anne1992276}. A $^{58}$Ni$^{26+}$ primary beam, with an average intensity of 3.7 e$\mu$A, was accelerated to 74.5 MeV/nucleon and fragmented on a natural Ni target, 200 $\mu$m thick. The LISE3 separator \cite{Anne1992276} was used to select the fragments, which were implanted at a rate of approximately 200 ions/s into a Double-Sided Silicon Strip Detector (DSSSD) of 300 $\mu$m thickness. The DSSSD had 16 X and 16 Y strips with a pitch of 3 mm, defining 256 pixels. Two parallel electronic chains were used having different gains to detect both the implanted heavy-ions and subsequent charged-particle (betas and protons) decays. The DSSSD was surrounded by four EXOGAM Ge clovers \cite{exogam} used to detect the $\beta$-delayed $\gamma$ rays.

The experiment was focused on the study of some $T_z=\text{-}2$ proton-rich nuclei. The energy of the $^{58}$Ni beam was optimized to implant $^{56}$Zn close to the middle of the DSSSD. Data were also taken by optimizing on $^{48}$Fe to increase the statistics for this ion. Further data were also recorded by focusing on the $T_z=\text{-}1$, $^{58}$Zn nucleus, of astrophysics interest since it constitutes a waiting point in the $rp$-process. The results from this dataset were used for comparison with a previous experiment and to estimate the $\beta$ detection efficiency in the DSSSD.

\begin{figure}[!ht]
  \centering
	\includegraphics[width=1\columnwidth,trim={0 0.2cm 3.1cm 5.5cm},clip]{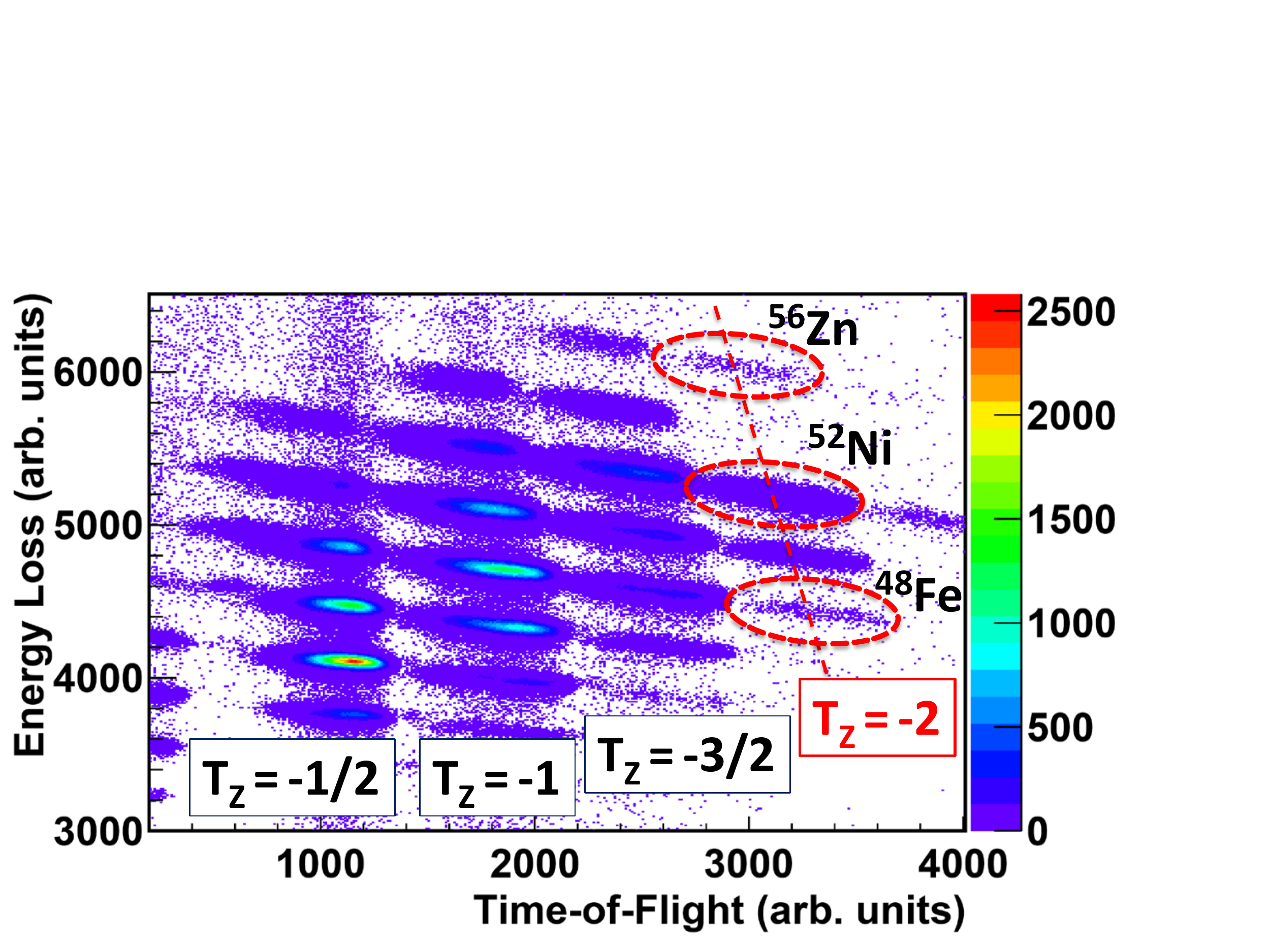}
	\vspace{-5.5 mm}
 	\caption{$\Delta E$ versus ToF identification plot for the dataset optimized to implant $^{56}$Zn close to the middle of the DSSSD. The positions of the $^{48}$Fe, $^{52}$Ni and $^{56}$Zn implants are shown.}
  \label{ID1plot}
\end{figure}

\begin{figure}[!ht]
  \centering
	\includegraphics[width=1\columnwidth,trim={0 0.2cm 3.0cm 5.6cm},clip]{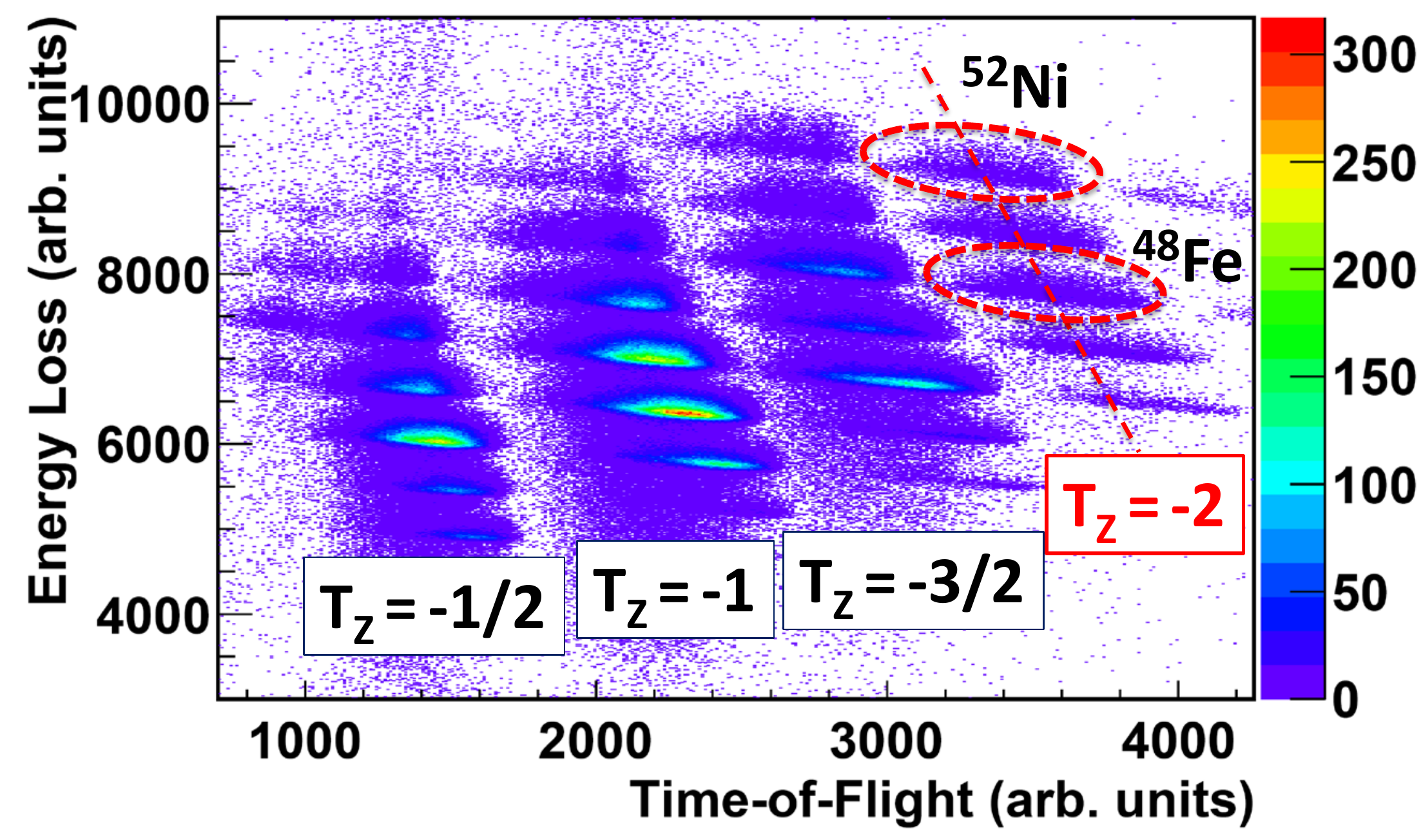}
	\vspace{-5.5 mm}
 	\caption{$\Delta E$ versus ToF identification plot for the dataset optimized for $^{48}$Fe. The positions of $^{48}$Fe and $^{52}$Ni are shown.}
  \label{ID2plot}
  \vspace{-5.0 mm}
\end{figure}

The Time-of-Flight (ToF) of the selected ions was defined as the time difference between the cyclotron radio-frequency and the signal they generated in a silicon $\Delta E$ detector located 28 cm upstream from the DSSSD. Simultaneous signals from both the $\Delta E$ detector and the DSSSD defined an implantation event. The implanted ions were identified by combining the energy loss signal in the $\Delta E$ detector and the ToF. An example of the two-dimensional identification matrix obtained for the setting focused on $^{56}$Zn is shown in Fig. \ref{ID1plot}, where the positions of the $T_z=\text{-}2$ nuclei $^{48}$Fe, $^{52}$Ni and $^{56}$Zn are indicated. The identification matrix obtained for the dataset optimized for $^{48}$Fe is shown in Fig. \ref{ID2plot}, where the positions of $^{48}$Fe and $^{52}$Ni are indicated. A signal above threshold (typically 50-90 keV) in the DSSSD and no coincident signal in the $\Delta E$ detector defined a decay event.

\section{\label{analysis}Data analysis}

The present section describes the analysis of the data related to the $\beta$ decays of $^{48}$Fe, $^{52}$Ni and $^{56}$Zn. Here we also expand on some of the details already presented in Ref. \cite{PhysRevLett.112.222501}. The $^{56}$Zn case is used as an example for the discussion. The $^{48}$Fe and $^{52}$Ni cases were analyzed following the same procedure; differences are also discussed.

The $^{48}$Fe, $^{52}$Ni or $^{56}$Zn ions were selected by setting gates off-line on the $\Delta E$-ToF matrix (see Figs. \ref{ID1plot} and \ref{ID2plot}). The total numbers of implanted nuclei detected, $N_{imp}$, were: $^{48}$Fe: 5.0x10$^{4}$, $^{52}$Ni: 5.3x10$^{5}$, $^{56}$Zn: 8.9x10$^{3}$.

\subsection{\label{correl}Time correlations}

In decay spectroscopy experiments performed with a continuous beam, no unequivocal correlation between a given implantation event and its corresponding decay event can be established \cite{Dossat200718}. A widely-used approach \cite{Dossat200718,PhysRevLett.112.222501,PhysRevC.91.014301} is to construct the correlations in time between all implantations and decay events. To this end, the 256 pixels of the DSSSD can be used as independent detectors. The $\it{correlation~time}$ $t_{corr}$ is defined as the time difference between a decay event in a given pixel of the DSSSD and any implantation signal that occurred before or after it in the same pixel that satisfied the conditions required to identify the nuclear species. The correlation condition is $|t_{corr}|=|t_{decay} - t_{implant}|<\mathcal{T}$, where $\mathcal{T}$ is a chosen period of time. This means that a given decay event will be correlated with all the implantations happening in the same pixel within the time $\mathcal{T}$, and vice versa. There will be a true correlation only when that decay event belongs to the implantation event under consideration. Otherwise correlations will be random because the decay event will be correlated artificially with an implantation happening before or after the implantation to which it really belongs.

This procedure ensures that the true correlations are taken into account, at the price of including many random correlations. The correlation-time spectrum and the energy spectra resulting from this method will therefore contain both true and random correlations.

\begin{figure}[!b]
  \centering
  \includegraphics[width=1\columnwidth]{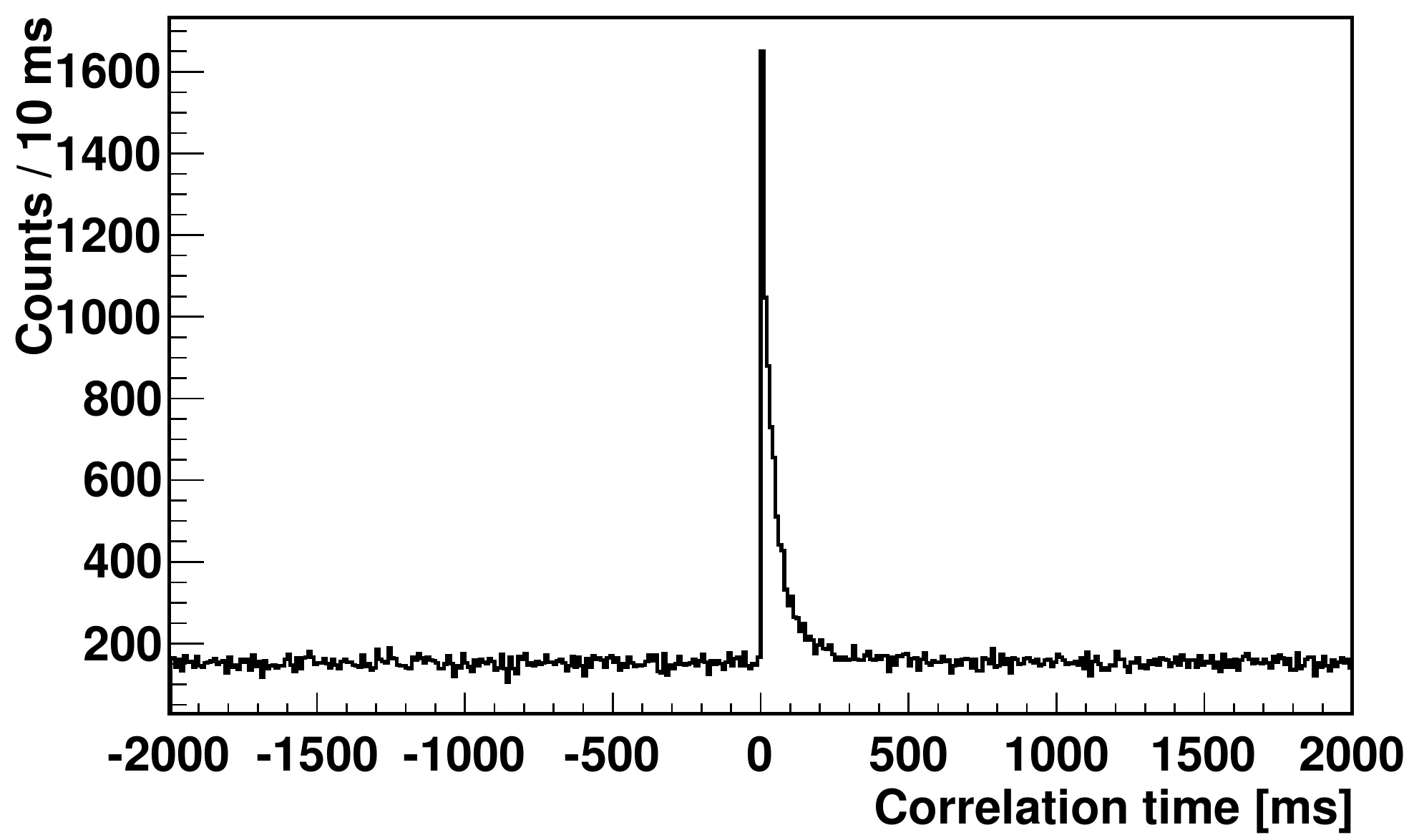}
  \vspace{-5.0 mm}
	\caption{Spectrum of the time correlations between each decay event and all of the $^{56}$Zn implants (see text).}
	\label{T56all}
\end{figure}

In the correlation-time spectrum the random events will form a flat background which can easily be taken into account, while the true correlations will form the typical exponential decay curve. As an example, the correlation-time spectrum for $^{56}$Zn including all the decays (betas and protons) is shown in Fig. \ref{T56all}.

Besides the truly-correlated decays, the energy spectra will contain decay events coming from either the same ion but a different implantation event, or from different nuclei. These randoms have to be carefully removed. The background subtraction procedure is described in Sections \ref{DSSSD} and \ref{gamma} for the charged-particle and $\gamma$-ray energy spectra, respectively.

We decided to correlate decays and implants in a period $\mathcal{T}$ = 50 s, i.e., in the interval [-50 s, +50 s]. This is different from Ref. \cite{Dossat200718} where the correlations were formed in the interval [0, 5000 ms]. There are two motivations for this choice. The $\it{correlations~in~the~past}$, [-50 s, 0], clearly have no physical meaning but they are useful in the background subtraction procedure (see Section \ref{DSSSD}). The choice of a huge period $\mathcal{T}$ allows one to check the shape of the background, which is expected to be flat (see Section \ref{bg}).

\subsection{\label{T}Determination of the half-life}

\subsubsection{\label{Tfit}Half-life fits}

For each nucleus of interest, the correlation-time spectrum can be constructed as explained in Section \ref{correl}. This spectrum includes all the decays (betas and protons) correlated with a given implant. Thus all the possible decays, from the parent nucleus or its daughters, have to be taken into account in the fit of the half-life $T_{1/2}$ with the Bateman equations \cite{Bateman1910}.

The detection efficiency for a particular decay event will depend strongly on the decay mode of that event, i.e., either by $\beta$-delayed proton or by $\beta$ emission alone. Indeed the detection efficiency of the DSSSD is very different for protons (close to 100\%) and for $\beta$ particles (\mbox{$\epsilon_{\beta}=13.6(6)$ \%}), see Section \ref{eff}. Therefore for nuclei where the decay mode is only by $\beta$ emission, the integration of the parent contribution and descendants will depend on the rather limited $\beta$ detection efficiency. 

$^{48}$Fe, $^{52}$Ni and $^{56}$Zn all emit $\beta$-delayed protons. Thus an additional condition can be imposed to select only the proton decays. This is achieved by setting an energy threshold in the DSSSD spectrum (Section \ref{DSSSD}) which removes the pure $\beta$ decays while keeping the $\beta$-delayed protons, and selecting correlated implants for the desired nuclear species. The advantage is that the daughter activity (where no protons are present) is removed from the correlation-time spectrum. In this way the half-life can be fitted simply by using the parent activity $A(t)$:
\begin{equation}
  A(t)=\lambda N_p~exp(-\lambda~t) \,,
  \label{Eq1}
\end{equation}
where \mbox{$\lambda = ln2/T_{1/2}$} is the radioactive decay constant and $N_p$ is the total number of proton emissions. Thus the integration of the parent contribution does not rely on $\epsilon_{\beta}$ and directly yields $N_p$. The background is fixed by a fit done on the left of the peak using a linear function.

\begin{figure}[!t]
  \centering
  \includegraphics[width=1.\columnwidth]{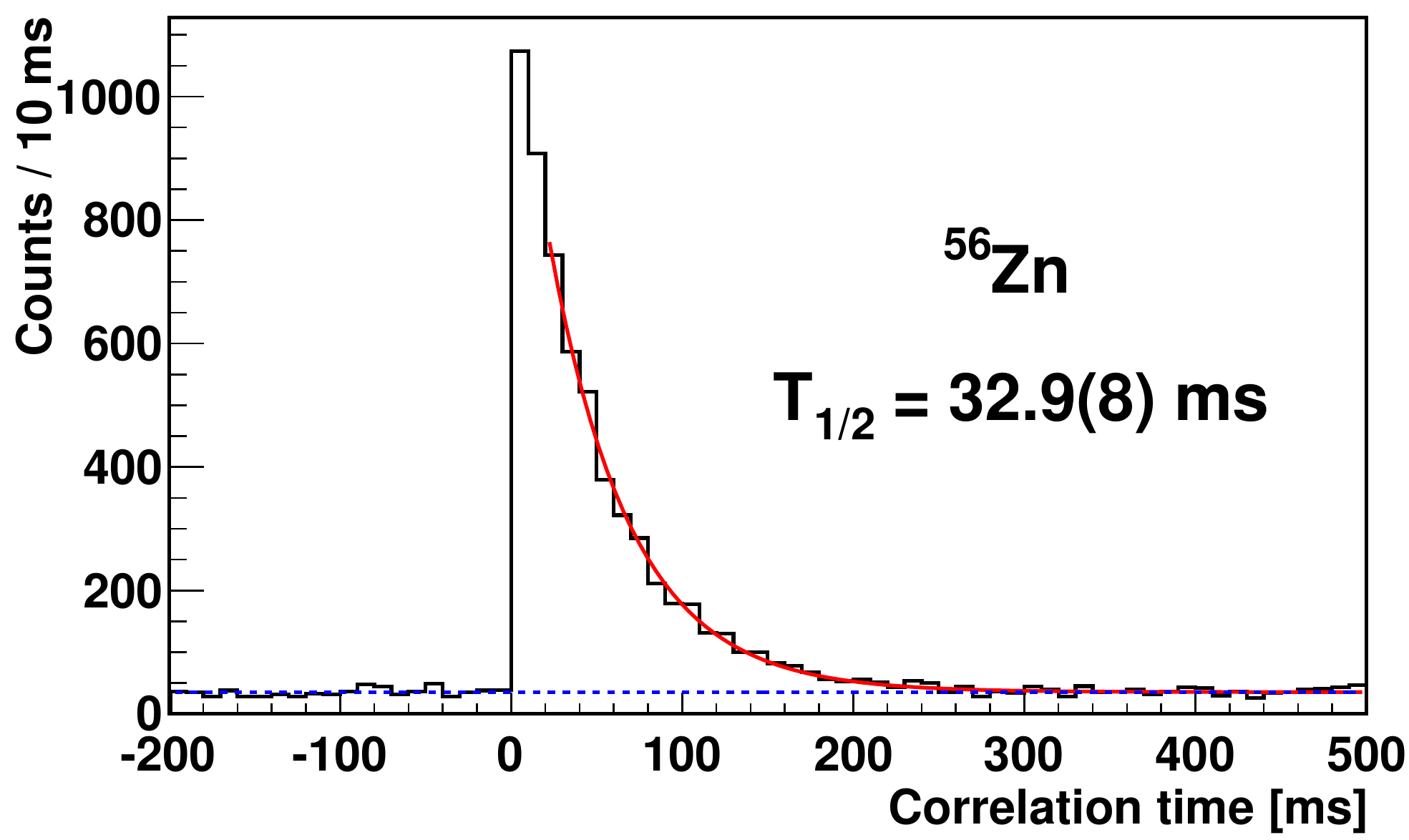}
  \vspace{-5.0 mm}
	\caption{Spectrum of the time correlations between each proton decay (DSSSD energy $>$ 0.8 MeV) and all $^{56}$Zn implants.}
	\label{T56p}
  \vspace{-5.0 mm}
\end{figure}

Fig. \ref{T56p} shows the correlation-time spectrum for $^{56}$Zn correlated with the proton decays alone (DSSSD energy threshold above 0.8 MeV). The half-life is determined by a least squares fit to the data with the function of Eq. \ref{Eq1}. A value of $T_{1/2}=32.9(8)$ ms is obtained for $^{56}$Zn \cite{PhysRevLett.112.222501}. A fit using a maximum likelihood minimization method gives a value of 32.8(8) ms. The result is not affected by the choice of the range over which the fit was made. Changing the range from 10 to 1000 half-lives (50 s) made no difference to the results.

For comparison, if we use the Bateman equations \cite{Bateman1910} to fit the $^{56}$Zn correlation-time spectrum containing all the decay modes (Fig. \ref{T56all}), we have to include: the $\beta$ decay of $^{56}$Zn to $^{56}$Cu, which partially goes to the ground state of $^{56}$Cu and partially undergoes proton emission to $^{55}$Ni; the $\beta$ decay of $^{56}$Cu to $^{56}$Ni; and the $\beta$ decay of $^{55}$Ni to $^{55}$Co. As expected, we get a similar value, 31.2(11) ms.

We adopt the first method because it is more precise, it relies only on our data and it allows us to extract $N_p$ independently of the $\beta$ detection efficiency.

\subsubsection{\label{bg}Background from the random correlations}

Various methods can be used to deal with the random-correlation background. As in Ref. \cite{Dossat200718}, we included a flat background in the half-life fit. This is a well-established method to be used in simple cases such as the present one, where the beam is continuous. If instead one subtracts the background before fitting, there is an additional source of uncertainty when fitting the final spectrum because of the larger fluctuations introduced by the subtraction.

Alternatively, a method that is very useful when the beam is pulsed has been developed in Refs. \cite{Molina2011,PhysRevC.91.014301}. There, a correlation-time spectrum for the background is constructed by performing the time correlations in the $opposite~pixel$, i.e., between decays in the pixel [$i$,$j$] and implants in the pixel [$j$,$i$] for $i \neq j$. The background spectrum is then subtracted from the main correlation-time spectrum. By applying this method in our case we get consistent results but a larger uncertainty on $T_{1/2}$ caused by the subtraction. Besides, the background spectrum introduces many fluctuations because of its low statistics. Indeed in our case most of the ions produced were implanted in the same region of the DSSSD as $^{56}$Zn, thus the opposite pixel can have few events when far from the implantation region. This is not the case for the nuclei studied in Refs. \cite{Molina2011,PhysRevC.91.014301}.

In order to increase the statistics of the background spectrum we have also considered an $extended$ background spectrum created by correlating the decays in the pixel [$i$,$j$] with the implants in all the pixels of the DSSSD except the 8 pixels surrounding the pixel [$i$,$j$] and the pixel [$i$,$j$] itself. In this way the statistics of the background spectrum are indeed improved, however when the beam is continuous we still prefer the first method because it avoids the introduction of additional fluctuations due to the background subtraction procedure.

In summary, as expected the three methods give consistent results. The choice of the method depends on the details of the experiment carried out (beam properties and implantation pattern). 

Finally, in Ref. \cite{Molina2011} a peculiar effect has been observed. Ideally the profile of the random-correlation background is expected to be a constant. However, there can be situations where the background profile is not flat, but has a shape. The shape can be affected, e.g., by pulsing of the beam, or by the limited duration of the runs. In the latter case the background assumes a triangular shape centred at $T_{1/2}=0$ \cite{Molina2011}. This effect is more pronounced as the runs become shorter and it may affect the determination of $T_{1/2}$. The effect and the procedure to correct for it will be described in detail elsewhere.

Here we performed the time correlations over a huge period $\mathcal{T}$ to study the background profile. In the present case the duration of the runs is large enough that the background profile is not affected. We verified that the results for the half-life fit are the same both with and without applying the correction for this effect.

\subsection{\label{eff}DSSSD detection efficiency}

A DSSSD detection efficiency of 100\% has been assumed for both implants and $\beta$-delayed protons. A rather low efficiency value is expected for the $\beta$ particles, also reflected in the $\beta$-delayed $\gamma$ emission, because of the combined effect of the small energy loss of the betas in the DSSSD detector (about 100-200 keV) and the electronic threshold. In contrast, the $\beta$-delayed proton emission yields a much higher signal in the DSSSD so that the proton detection efficiency is not affected by the threshold. If the implantation occurs in the middle of the DSSSD, the proton efficiency is close to 100\%, as explained below.

A Monte Carlo simulation of the detection efficiency for protons is shown as a function of their energy in Fig. 5 of Ref. \cite{Dossat200718}, where different implantation depths in a 300 $\mu$m thick DSSSD detector are considered. It was found that the detection efficiency was symmetric with respect to the centre of the detector. If the $^{56}$Zn ions are implanted in the centre of the DSSSD (150 $\mu$m), the efficiency for protons of energy up to 3.5 MeV is 100\%. We simulated the implantation profile of the $^{56}$Zn ions in the DSSSD in realistic conditions using the simulation code LISE \cite{Tarasov20084657}, obtaining a distribution centred in the DSSSD and having a width of 30 $\mu$m FWHM. From previous similar works, such as Ref. \cite{Dossat200718}, the GANIL experts indicate that the differences in the implantation depth between the LISE simulations and the measurements are of the order of 10 $\mu$m. If we adopt a reasonable systematic error of 20 $\mu$m in the implantation depth, then the detection efficiency is 100\% for all the protons with energy $\leq$3.0 MeV. Only for the 3.45 MeV proton seen in the decay of $^{52}$Ni is the efficiency slightly lower ($\geq$ 98.5\%).

The $\beta$ detection efficiency of the DSSSD is determined by means of known $\beta\gamma$ emitters, i.e., the $T_z=\text{-}1$ nuclei populated in the dataset focused on $^{58}$Zn \cite{Orrigo:2014gla}. As distinct from the $T_z=\text{-}2$ nuclei under study, $\beta$-delayed proton emission is not present in the decay of these $T_z=\text{-}1$ isotopes. Indeed their decay proceeds by either $\beta$-delayed $\gamma$ emission or by $\beta$ decay to the ground state, thus they can be used to estimate the $\beta$ detection efficiency from:
\begin{equation}
  \epsilon_{\beta} = \frac{N_{\beta}}{N_{imp}~(1-\tau)} \,,
  \label{Eq2}
\end{equation}
where $N_{\beta}$ is the number of $\beta$ decays obtained by integrating the parent activity from the correlation-time spectrum for these nuclei and $\tau$ is the dead time fraction, which varies from $\approx$12\% in the $^{58}$Zn dataset to $\approx$28\% in the $^{56}$Zn+$^{48}$Fe datasets. The number of implants $N_{imp}$ is obtained by selecting the ion of interest in the identification matrix (Section \ref{analysis}). We obtain $\epsilon_{\beta}=13.6(6)~\%$. Monte Carlo simulations of the emitted $\beta$ particles (done as in Section \ref{DSSSD} for the $\beta$-p events) showed that the differences in $\epsilon_{\beta}$ between $^{58}$Zn and the $T_z=\text{-}2$ nuclei under study ($^{48}$Fe, $^{52}$Ni and $^{56}$Zn) are of the order of 1\%, hence well inside the quoted uncertainty for $\epsilon_{\beta}$.

As explained in Section \ref{exp}, the data acquisition system was triggered by either an implantation event or by a decay event ($\beta$ or proton). The $\gamma$ rays were not included in the trigger. The $\gamma$ emissions were acquired in coincidence with the decay events, which are affected by the corresponding DSSSD efficiency. The $\beta$ efficiency is used to determine the absolute intensity $I_{\gamma}$ of each $\beta$-delayed $\gamma$ ray observed in the decays of $^{48}$Fe and $^{52}$Ni from:
\begin{equation}
  I_{\gamma}^{(\beta-\gamma)} = \frac{N_{\gamma}}{\epsilon_{\gamma}~\epsilon_{\beta}~N_{imp}~(1-\tau)} \,,
  \label{Eq3}
\end{equation}
where $N_{\gamma}$ represents the number of counts in the given $\gamma$ line and $\epsilon_{\gamma}$ is the $\gamma$ detection efficiency of the EXOGAM Ge clovers (see Section \ref{gamma}). For $^{56}$Zn, instead, the observed $\gamma$ lines are in coincidence with the proton emission, therefore their intensity does not depend on $\epsilon_{\beta}$:
\begin{equation}
  I_{\gamma}^{(p-\gamma)} = \frac{N_{\gamma}}{\epsilon_{\gamma}~N_{imp}~(1-\tau)} \,.
  \label{Eq4}
\end{equation}

\subsection{\label{Bp}Total proton-emission branching ratio}

The total proton branching ratio $B_p$ is determined by comparing the total number of protons, $N_p$, with the total number of implanted nuclei, $N_{imp}$, according to: 
\begin{equation}
  B_p = \frac{N_p}{N_{imp}~(1-\tau)} \,.
  \label{Eq5}
\end{equation}

$N_p$ is obtained, together with the half-life, from a fit of the correlation-time spectrum after selection of the proton decays (see Section \ref{Tfit}). The selection of the proton emission is achieved by setting an energy threshold in the DSSSD spectrum just below the first proton peak we identify. As discussed in Ref. \cite{Dossat200718}, the systematic error of this procedure can be estimated by repeating the determination of $B_p$ using a DSSSD threshold differing by $\pm$100 keV. Since the dead time fraction $\tau$ could change in the different measurements, we calculated a weighted-average dead time in the same way as in Ref. \cite{Dossat200718}. The dead time recorded in each run was weighted by the number of implants of a given ion in that run and divided by the total $N_{imp}$ of that isotope.

\subsection{\label{DSSSD}Analysis of the charged-particle spectrum}

A 300 $\mu$m thick DSSSD detector was used to detect both the implanted heavy-ions and the following charged-particle (betas and protons) decays (see Section \ref{exp}). The strips of the DSSSD were calibrated and aligned using a triple-$\alpha$-particle source and the peaks of known energy from the decay of $^{53}$Ni \cite{Dossat200718}. The DSSSD spectrum was obtained as the sum of the spectra from all the 256 pixels.

\begin{figure}[!h]
  \centering
  \includegraphics[width=1\columnwidth]{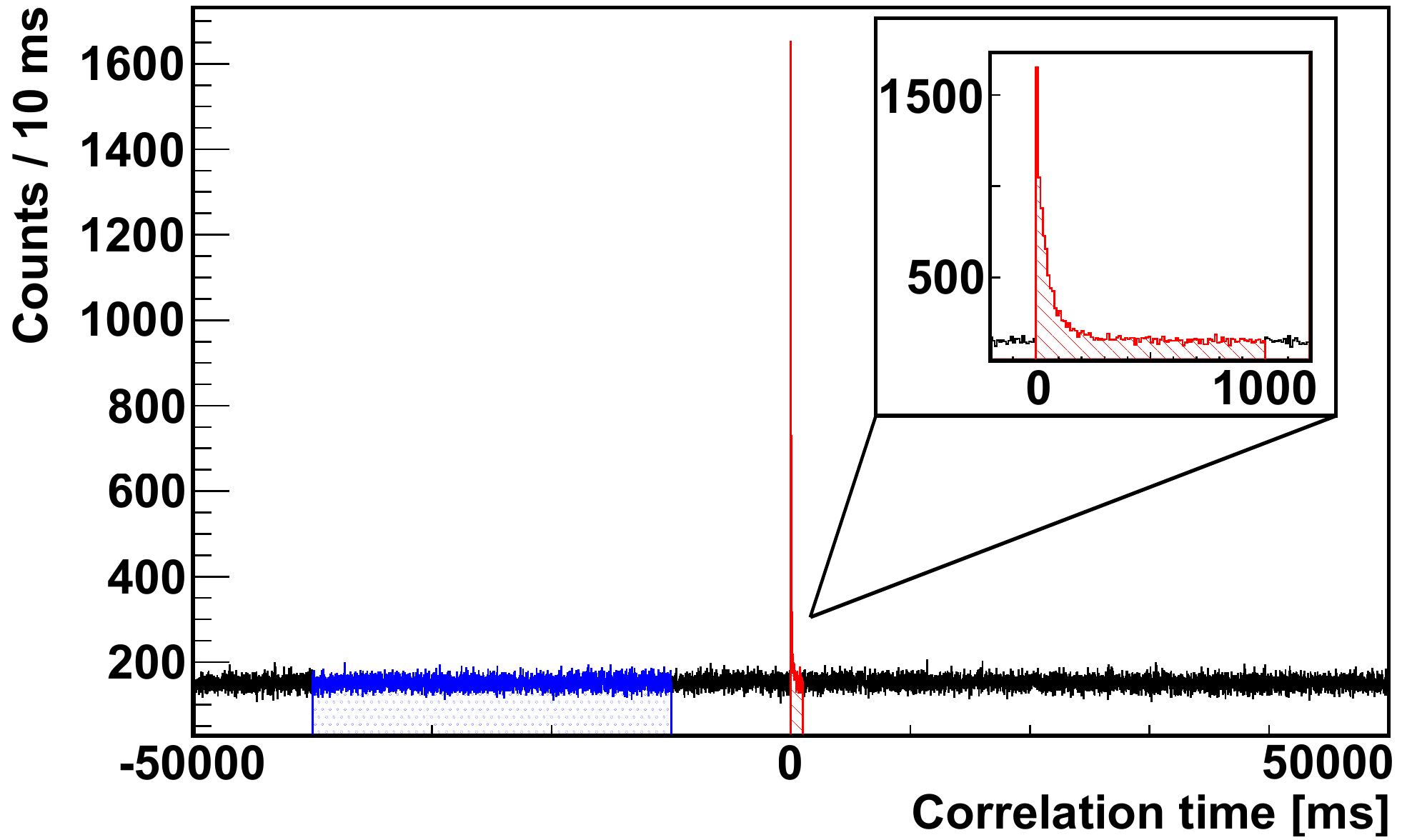}
  \vspace{-5.0 mm}
	\caption{Spectrum of the time correlations between each decay event and all the $^{56}$Zn implants. The 0 to 1 s and -40 to \mbox{-10 s} gates are indicated by the red-dashed and blue-dotted regions, respectively.}
	\label{T56forBG}
\end{figure}

\begin{figure}[!t]
	\begin{minipage}{1.0\linewidth}
    \centering
    \includegraphics[width=1\columnwidth]{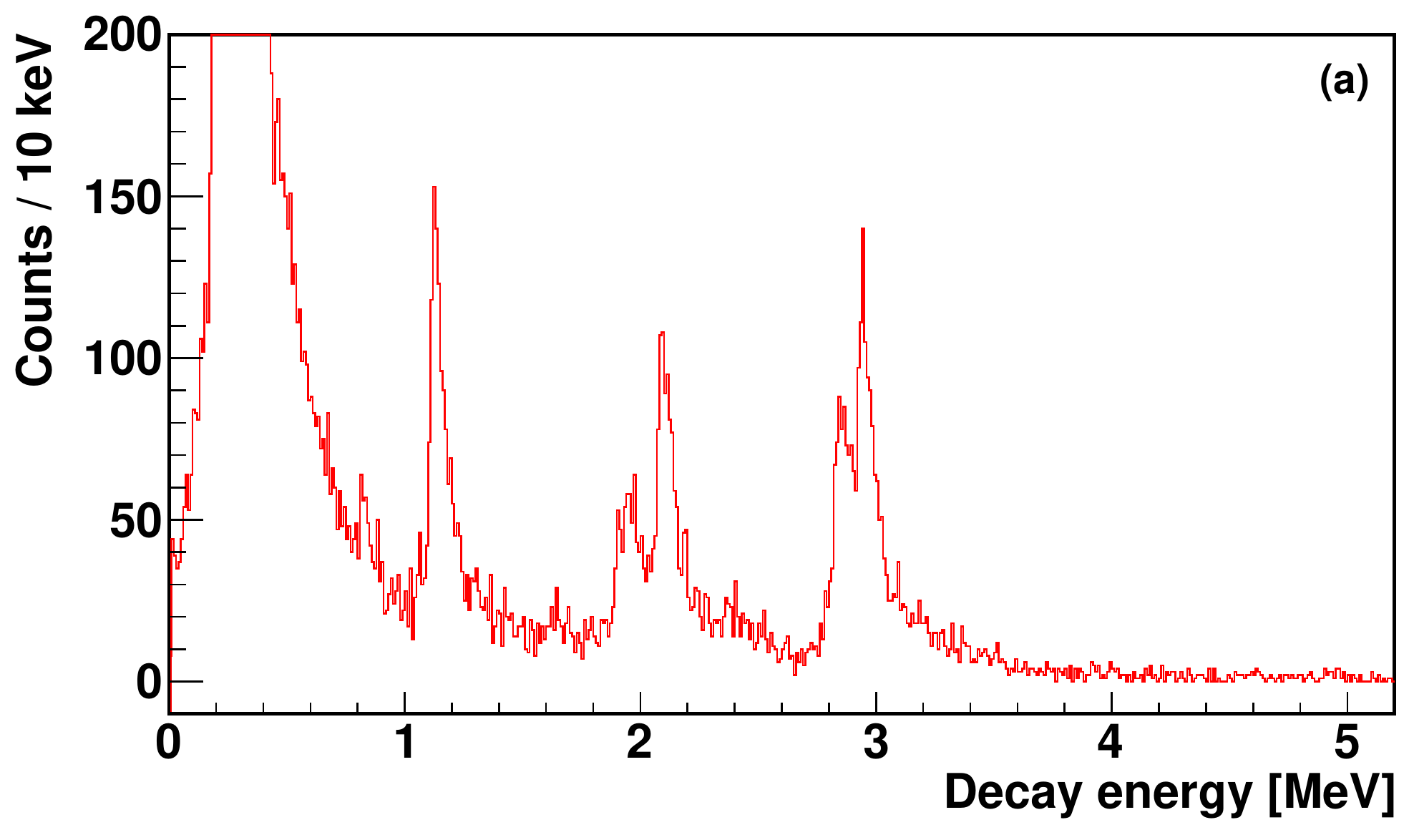}
	\end{minipage}
	\begin{minipage}{1.0\linewidth}
	  \centering
    \includegraphics[width=1\columnwidth]{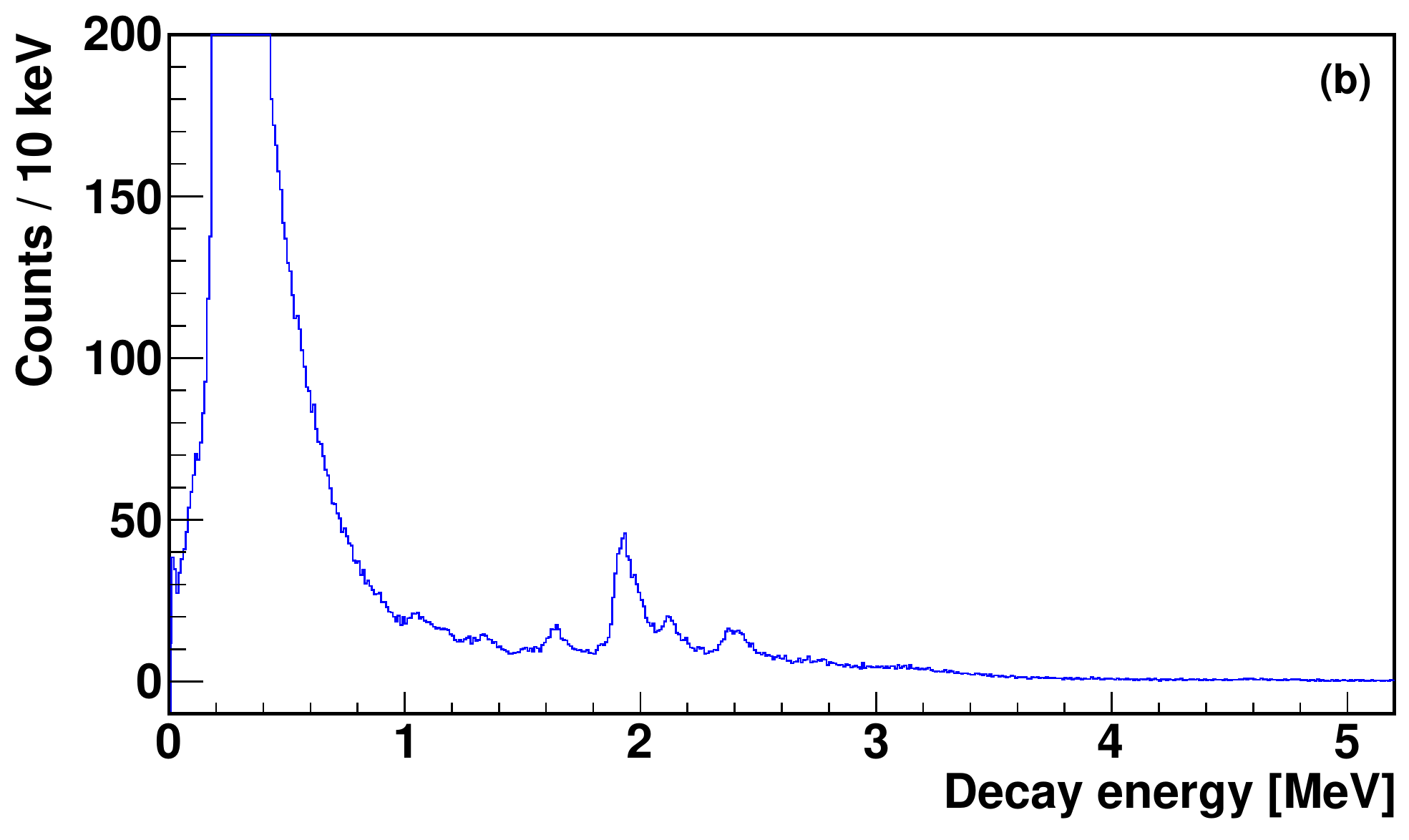}
	\end{minipage}
	\begin{minipage}{1.0\linewidth}
	  \centering
    \includegraphics[width=1\columnwidth]{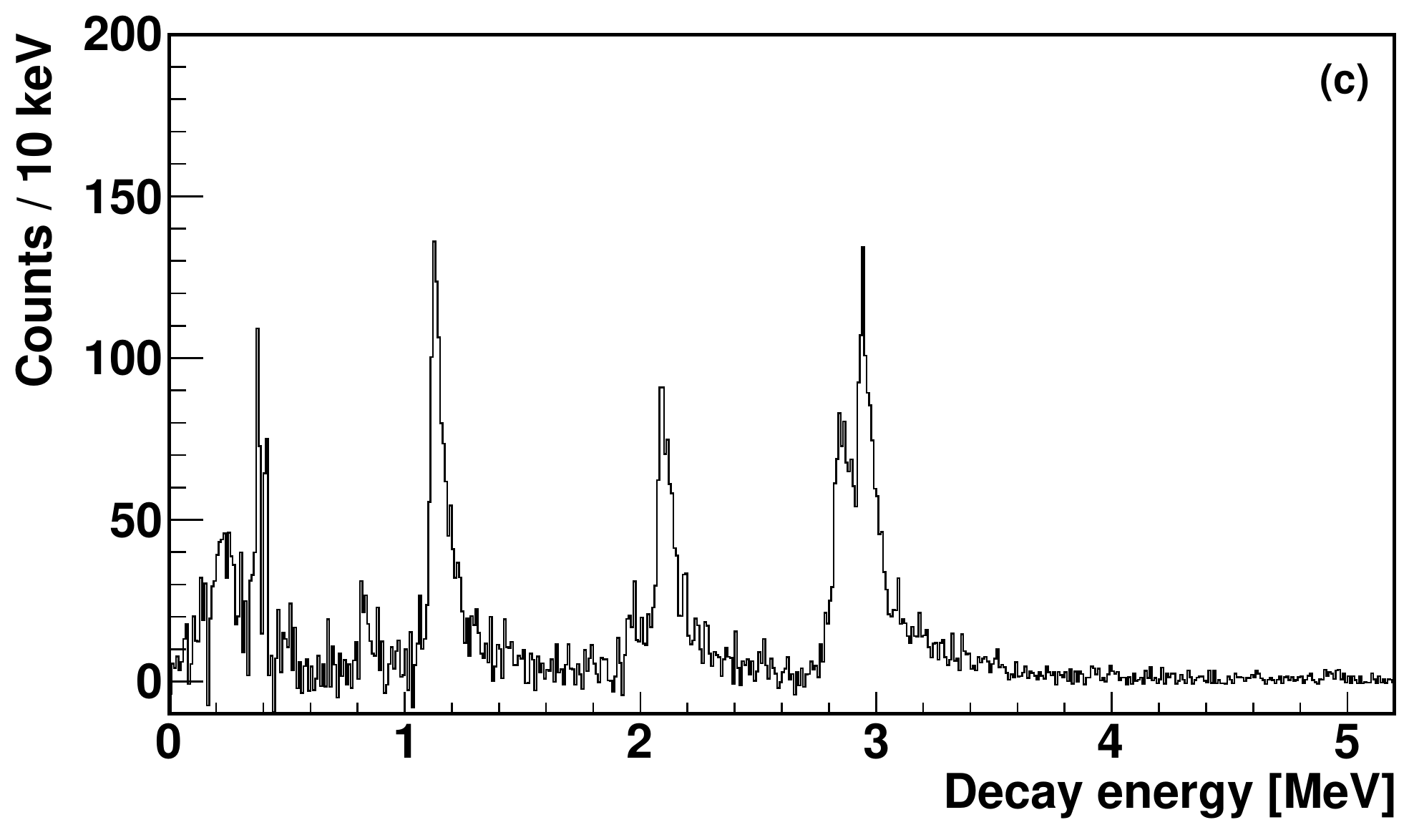}
	\end{minipage}
	\caption{DSSSD charged-particle spectra for decay events correlated with $^{56}$Zn implants. The spectrum (a) corresponds to time correlations from 0 to 1 s (red-dashed zone in Fig. \ref{T56forBG}) and contains both true and random correlations. The background spectrum (b) is created for time correlations from -40 to -10 s (blue-dotted zone in Fig. \ref{T56forBG}), normalized by the time interval used and contains only randoms. Peaks from the $^{53}$Ni contaminant ($T_{1/2}$ = 55.2(7) ms) \cite{Dossat200718} are seen between 1 and 2.5 MeV. The removal of the randoms is achieved in spectrum (c), obtained by subtracting the spectrum (b) from (a).}
	\label{dsssdbg}
  \vspace{-5.0 mm}
\end{figure}

The charged-particle spectrum measured in the DSSSD for decays associated with implants of a given nuclear species was formed as follows (the figures shown as examples are related to $^{56}$Zn). A DSSSD spectrum was created by selecting time correlations from 0 to 1 s (red-dashed zone in Fig. \ref{T56forBG}). This DSSSD spectrum, shown in Fig. \ref{dsssdbg}a, contains both true and random correlations. A background DSSSD spectrum, containing only randoms and shown in Fig. \ref{dsssdbg}b, was formed by setting a gate from -40 to -10 s on the time correlation spectrum (selecting the blue-dotted zone in Fig. \ref{T56forBG}). The two spectra were then normalized to the time interval used and the spectrum derived from Fig. \ref{dsssdbg}b was subtracted from the spectrum in Fig. \ref{dsssdbg}a. The resulting DSSSD spectrum associated with $^{56}$Zn is shown in Fig. \ref{dsssdbg}c.

Our procedure is similar to that of Ref. \cite{Dossat200718}, but there the gate for the background spectrum was chosen from 1 to 2 s. We chose a larger time gate, 30 s, to increase the statistics of the background spectrum. The advantage is a reduction in the fluctuations coming from the subtraction of the two spectra. This is important in cases such as $^{56}$Zn where the number of counts is limited. Moreover, we use the $\it{correlations~in~the~past}$ to define a time interval on the left of the peak, which ensures that only randoms are included in the background spectrum.

In general, a DSSSD spectrum shows a bump at low energy, due to the detection of $\beta$ particles. The discrete peaks visible above this bump are interpreted as being due to $\beta$-delayed proton emission. The proportion of $\beta$-delayed protons to $\beta$ decay depends on the nucleus under study and it is reflected in the value of $B_p$ (Eq. \ref{Eq5}). For example, for $^{56}$Zn the strength in Fig. \ref{dsssdbg}c is dominated by $\beta$-delayed proton emission and $B_p$ = 88.5(26) \% \cite{PhysRevLett.112.222501}.

\begin{figure}[!t]
	\begin{minipage}{1.0\linewidth}
    \centering
    \includegraphics[width=1\columnwidth]{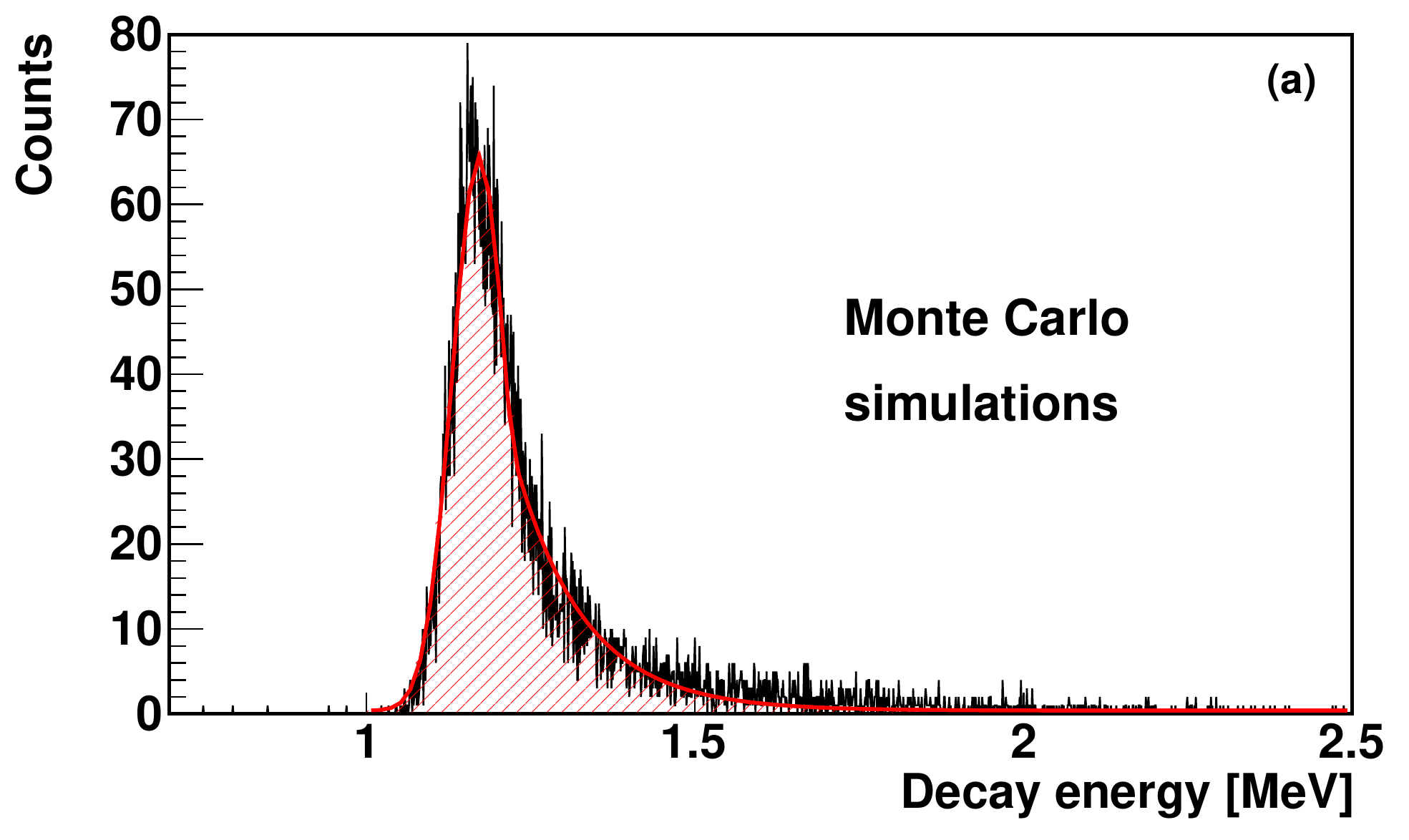}
	\end{minipage}
	\begin{minipage}{1.0\linewidth}
	  \centering
    \includegraphics[width=1\columnwidth]{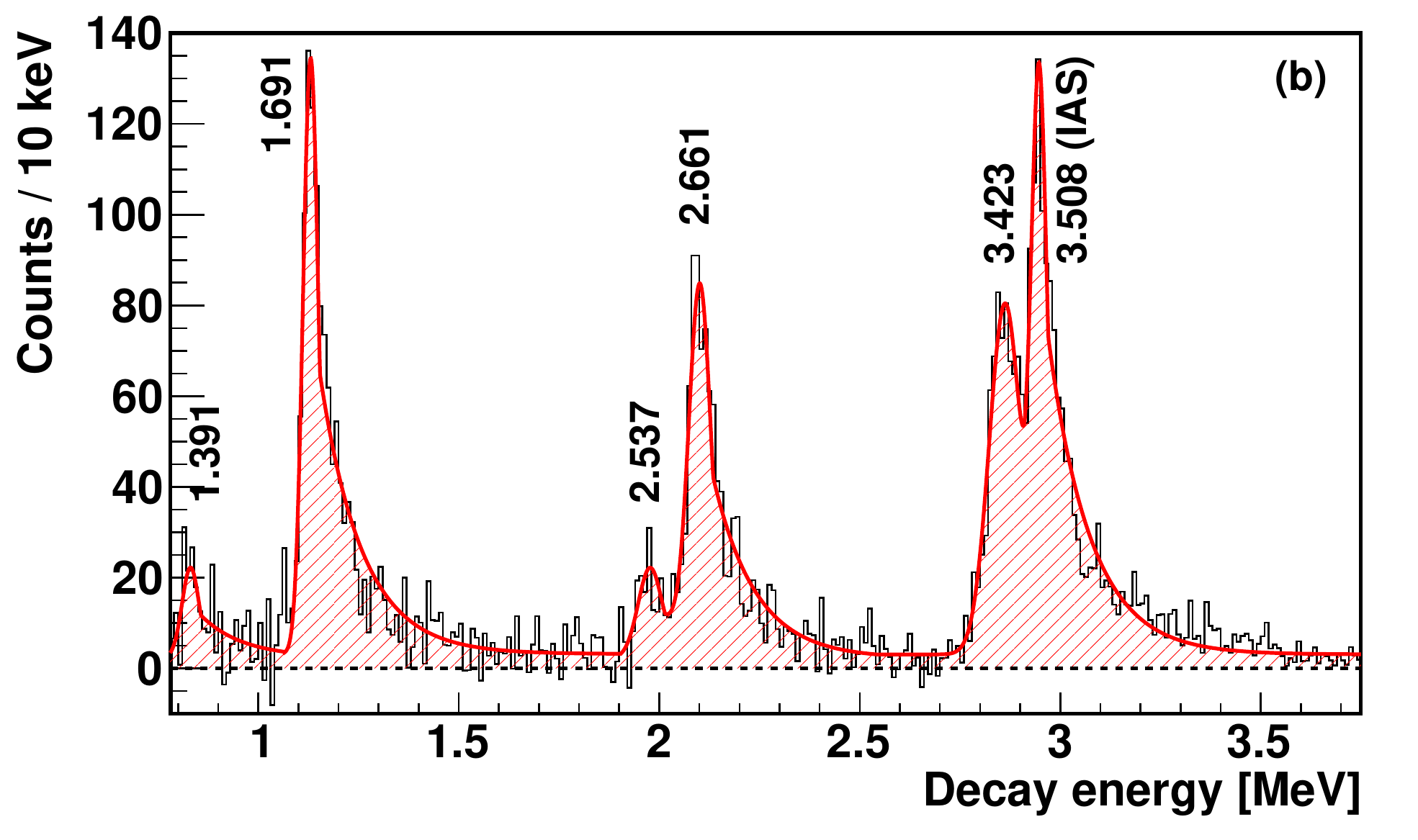}
	\end{minipage}
	\caption{(a) Monte Carlo simulations of $\beta$-p events with \mbox{$E_p=1.1$ MeV} and a $\beta$ end-point energy of 11.2 MeV. The peak is fitted by a Gaussian function having an exponential high-energy tail. (b) DSSSD charged-particle spectrum for decay events correlated with $^{56}$Zn implants. The lineshape of Fig. \ref{MC}a was used for the fit shown here in dashed-red.}
	\label{MC}
  \vspace{-5.0 mm}
\end{figure}

The DSSSD experimental energy resolution is 70 keV FWHM. The summing with the coincident $\beta$ particles also affects the lineshape of the peak. To study the lineshape we performed Monte Carlo simulations for a silicon DSSSD strip using the Geant4 code (version 4.9.6) \cite{geant2003,geant2006}. The radioactive sources (i.e., the implants) were located in an extended area in the middle of the detector, with an implantation profile obtained from LISE calculations \cite{Tarasov20084657}. Protons of a given energy $E_p$ were emitted at the same time as $\beta$ particles. The latter follow a distribution determined by the Fermi function ($\beta$-decay event generator), with an end-point energy corresponding to $Q_{\beta}-E_p-S_p$, where $S_p$ is the proton separation energy in the daughter nucleus. A widening of 70 keV FWHM was imposed to simulate the DSSSD experimental resolution.

As an example, Fig. \ref{MC}a shows the result of the simulation of the $\beta$-p decay with $E_p$ = 1.1 MeV (corresponding to the $^{56}$Cu level at 1.7 MeV, see Table \ref{table56}). A Gaussian with an exponential high-energy tail describes the lineshape to a good approximation. This result is an additional justification for the procedure widely used in Ref. \cite{Dossat200718}. The lineshape obtained from the simulations was also checked by us with the well isolated $^{57}$Zn proton peak at $E_p$ = 4.6 MeV. We used this lineshape to fit the experimental DSSSD spectra, by keeping the shape fixed (i.e., the Gaussian+exponential and the slope of the exponential) and fitting the parameters of the Gaussian (height, mean, width) to the experimental proton peaks. The result for $^{56}$Zn is shown in Fig. \ref{MC}b. The subroutine \cite{Jerome}, which allows the use of the same function in a more automated way, was used to cross-check the fit of $^{56}$Zn and to perform the fits for $^{48}$Fe and $^{52}$Ni, where many small peaks overlap.

The procedures adopted allow for a proper determination of the proton centre-of-mass energy $E_p$ and the number of counts $N_p$ in each proton peak $i$. The intensities $I_p^i$ of the proton peaks, which are important for the extraction of the $\beta$-decay strengths, are then given by:
 \begin{equation}
  I_p^i = \frac{N_p^i}{N_{imp}~(1-\tau)} \,.
  \label{Eq6}
\end{equation}
The excitation energy $E_X$ of each level $i$ populated in the daughter nucleus is obtained by adding $E_p$ and $S_p$. It should be noted that in this kind of experiment both the kinetic energy of the proton and the recoil energy are absorbed in the DSSSD.

\subsection{\label{gamma}Analysis of the gamma spectrum}

Four EXOGAM Ge clovers were used to detect the $\beta$-delayed $\gamma$ rays. Each clover comprised four Ge crystals, giving a total of 16 Ge crystals. Two parallel electronic chains with different amplifiers were used to detect $\gamma$ rays up to 2 MeV and $\gamma$ rays of higher energy. The first electronic chain was used in the analysis of all the $\gamma$ rays up to 2 MeV. In the spectra obtained by using the second electronic chain a problem was detected during the data analysis. As seen from known peaks, a regular deformation pattern was observed, i.e., a distortion of each peak in a interval of $\approx$60 keV. This affected the analysis of the higher-energy $\gamma$ rays, particularly for $^{48}$Fe because of the poor statistics.

\begin{figure}[!b]
  \centering
  \includegraphics[width=1\columnwidth]{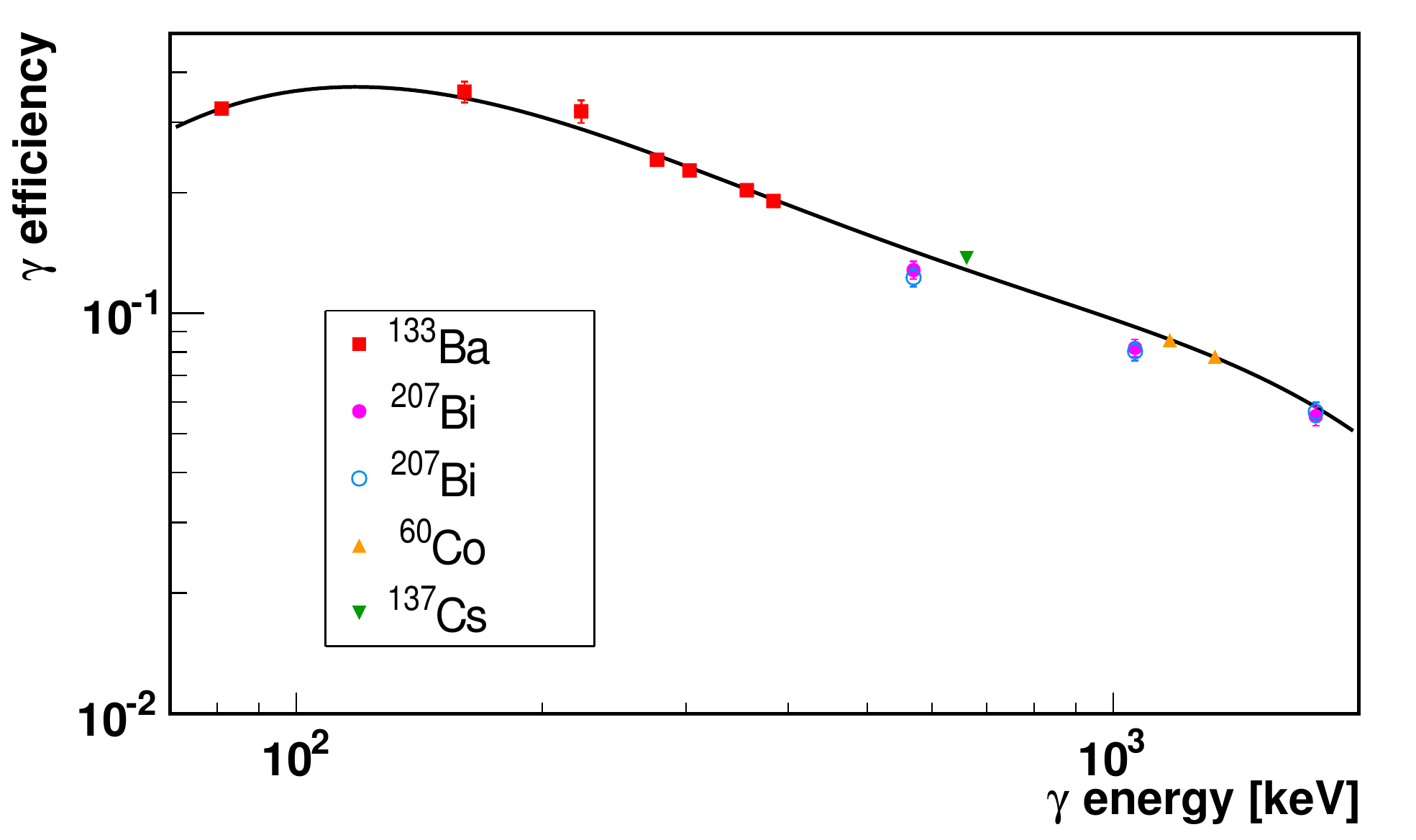}
  \vspace{-5.0 mm}
	\caption{Calibration of the efficiency for $\gamma$ detection. Data points from the $^{60}$Co, $^{137}$Cs, $^{133}$Ba and $^{207}$Bi sources are shown, together with a fit by the function from Ref. \cite{Hu1998121}.}
	\label{effg}
\end{figure}

The Ge crystals were calibrated in energy by using the $\gamma$ lines from $^{60}$Co, $^{137}$Cs and $^{152}$Eu sources. The alignment between the spectra measured in the 16 crystals was cross-checked and a summed spectrum was constructed. The energy resolution was 4 keV FWHM for the $^{60}$Co line at 1.33 MeV. The $\gamma$ detection efficiency was calibrated by using the $\gamma$ lines from the $^{60}$Co, $^{137}$Cs, $^{133}$Ba and $^{207}$Bi sources. An efficiency curve, shown in Fig. \ref{effg}, was fitted to the experimental data according to the efficiency energy dependence given in Ref. \cite{Hu1998121}. The $\gamma$ efficiency was $\approx$10\% at 1 MeV $\gamma$-ray energy.

\begin{figure}[!b]
	\begin{minipage}{1.0\linewidth}
    \centering
    \includegraphics[width=1\columnwidth]{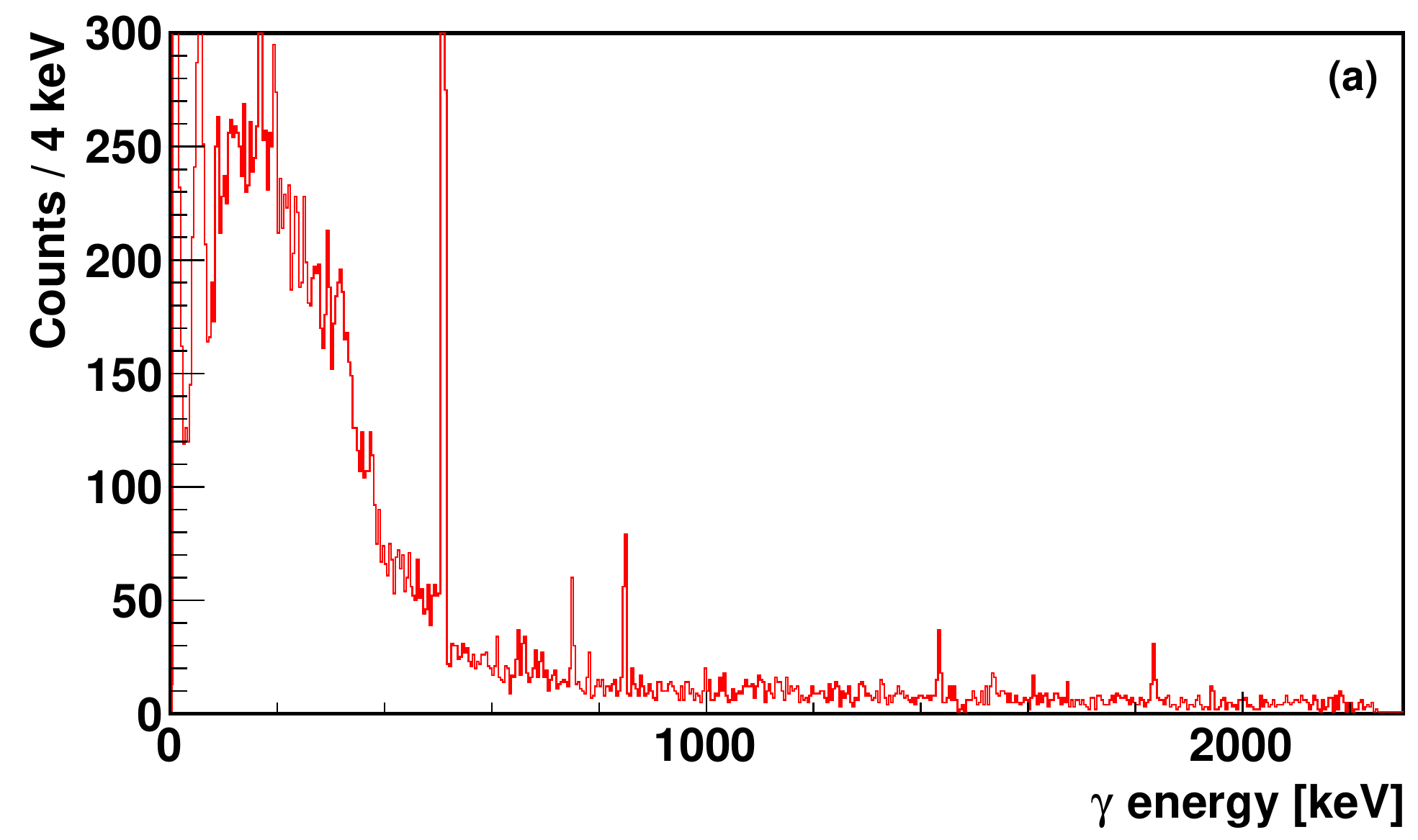}
	\end{minipage}
	\begin{minipage}{1.0\linewidth}
	  \centering
    \includegraphics[width=1\columnwidth]{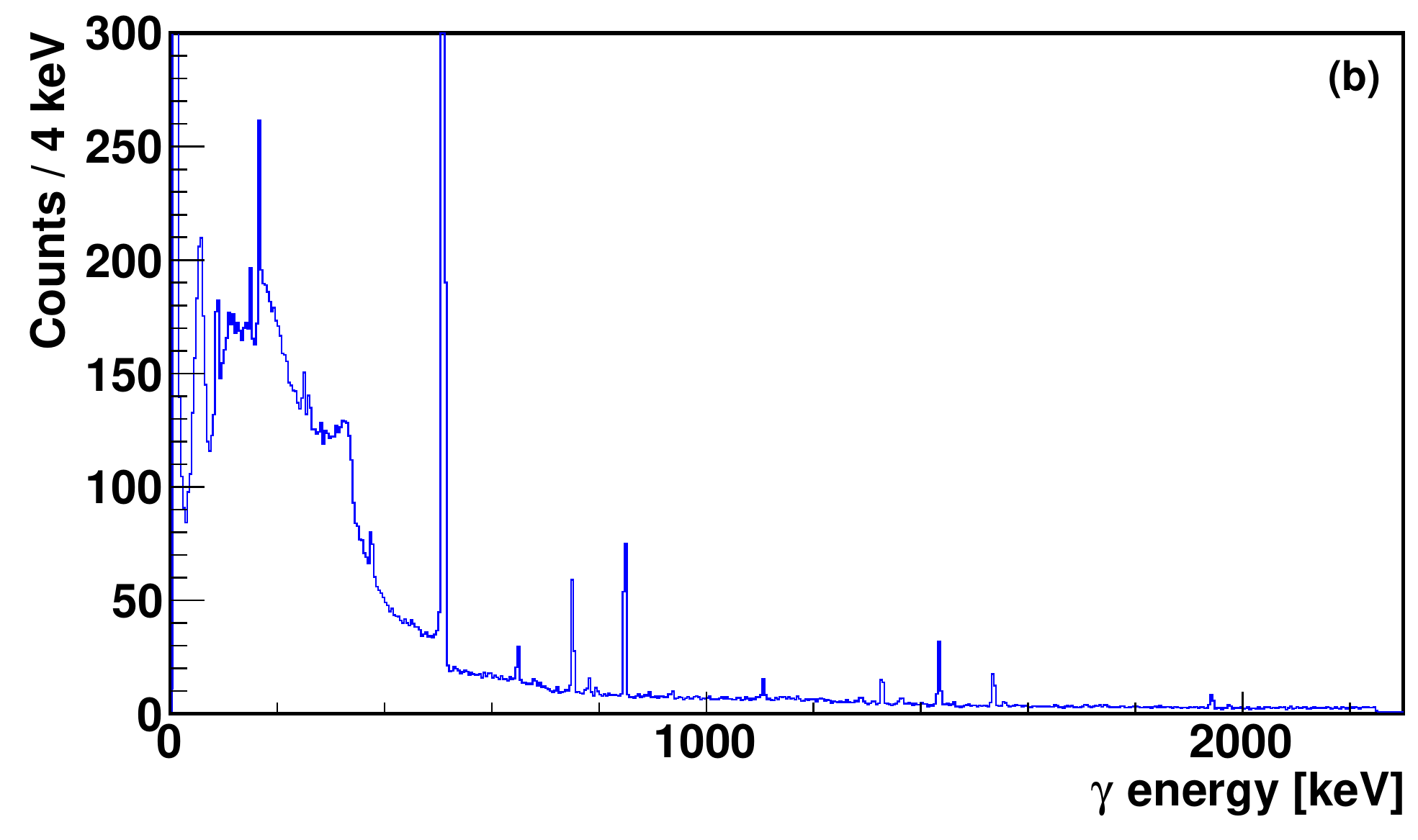}
	\end{minipage}
	\begin{minipage}{1.0\linewidth}
	  \centering
    \includegraphics[width=1\columnwidth]{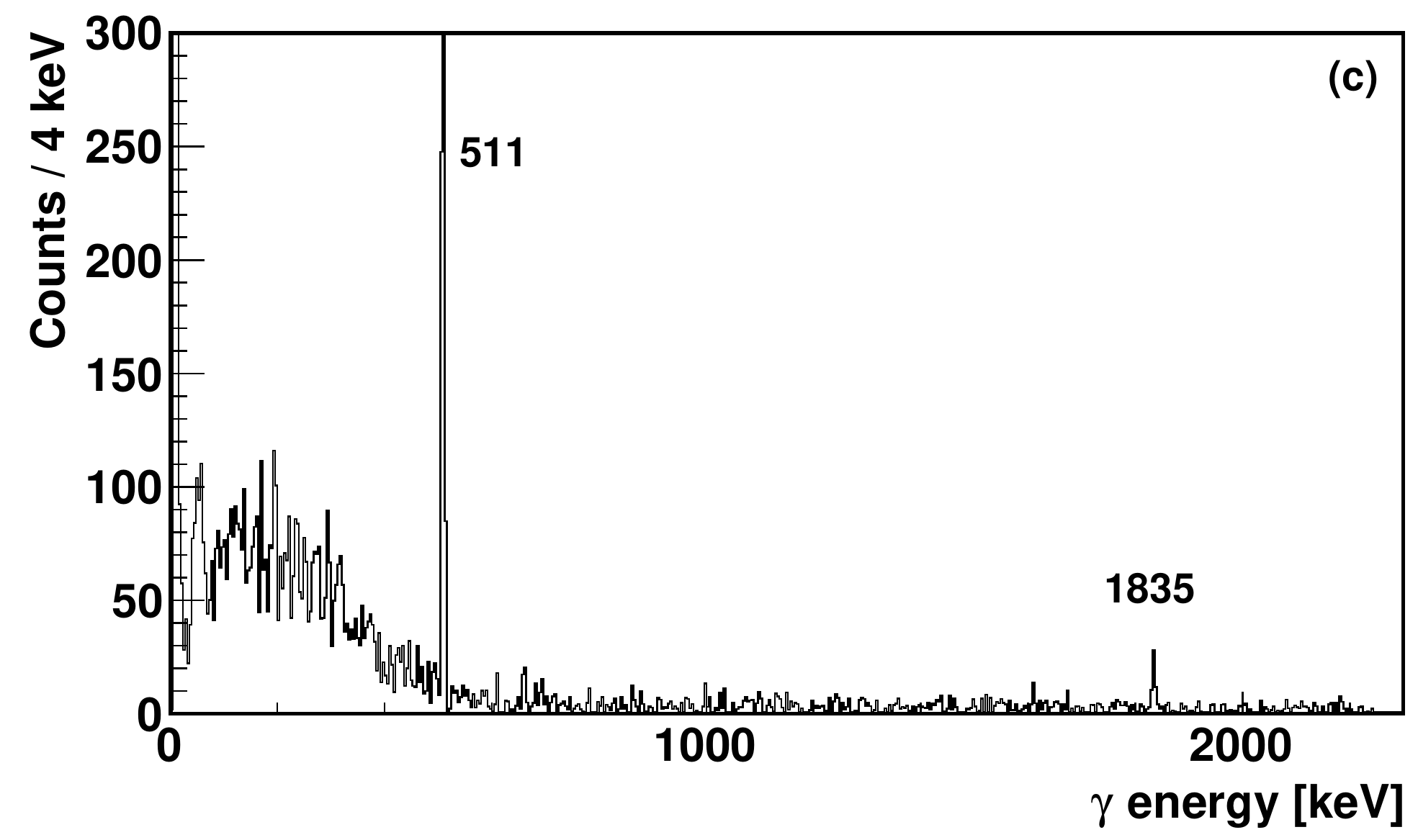}
	\end{minipage}
	\caption{$\gamma$-ray spectra for decay events correlated with $^{56}$Zn implants. The spectrum (a) corresponds to time correlations from 0 to 1 s (red-dashed zone in Fig. \ref{T56forBG}) and contains both true and random correlations. The background spectrum (b) is created for time correlations in the blue-dotted zone in Fig. \ref{T56forBG}, normalized by the time interval used and contains only randoms. The removal of the randoms is achieved by subtracting the spectrum (b) from (a) and it is shown in the spectrum (c). In addition to the 511 keV $\gamma$ line associated with the annihilation of the positrons emitted in the decay of $^{56}$Zn, a peak belonging to the daughter $^{56}$Cu is seen at 1835 keV \cite{PhysRevLett.112.222501}.}
	\label{gammabg}
\end{figure}

In order to create the $\gamma$ spectrum correlated with implants of a given nuclear species, one has first to define a $\beta$(-p)-$\gamma$ event. This is done by taking all the $\gamma$ events in coincidence with the $\beta$(-p) decays. Then the $\beta$(-p)-$\gamma$ events are correlated with the implantation signals, in the same way as was done in Section \ref{correl}. The background subtraction procedure is analogous to that adopted to create the DSSSD spectrum (Section \ref{DSSSD}). An example of the resulting $\gamma$ spectra is shown in Fig. \ref{gammabg} for $^{56}$Zn. A $\gamma$ spectrum was created by selecting ($\beta$(-p)-$\gamma$)-implant time correlations from 0 to 1 s, which contains both true and random correlations (Fig. \ref{gammabg}a). A background $\gamma$ spectrum, containing only randoms, was formed by setting a gate on the $\it{correlations~in~the~past}$ (Fig. \ref{gammabg}b). The two spectra were then normalized to the time interval used and the spectrum in Fig. \ref{gammabg}b was subtracted from that in Fig. \ref{gammabg}a to produce Fig. \ref{gammabg}c.

The $\gamma$ spectrum coincident with decays correlated with a given ion is then analyzed. The $\gamma$ lines of this spectrum are good candidates for $\beta$-delayed $\gamma$ transitions belonging to the ion of interest. The half-life associated with each $\gamma$ line is determined from the fit of a correlation-time spectrum gated on that line, after subtraction of a background correlation-time spectrum obtained by setting an energy gate on the two sides of the $\gamma$ line (mainly Compton background). This half-life is compared to the half-life of the nucleus. Such a procedure allowed the association of the 1835 keV $\gamma$ line with $^{56}$Zn \cite{PhysRevLett.112.222501}.

Gamma-proton coincidences are also obtained either by placing conditions on the proton peaks of the DSSSD spectrum or by putting gates on the $\gamma$ lines. In cases such as $^{56}$Zn, where most of the $\gamma$ decay is followed by a proton emission, the $\gamma$-proton coincidences allow for the identification of weak $\gamma$ lines and further confirm that these transitions belong to the decay of $^{56}$Zn.

Each $\gamma$ line was fitted by a Gaussian with a linear background to extract the energy centroid $E_{\gamma}$ and number of counts $N_{\gamma}$. The intensity of the line, $I_{\gamma}$, was obtained according to either Eq. \ref{Eq3} or \ref{Eq4} depending on the case.

\subsection{\label{strength}Determination of $B$(F) and $B$(GT)}

For the even-even $T_z$ = -2 nuclei discussed here, two kinds of state are expected to be populated in $\beta$ decay: the $T$ = 2, $J^{\pi}$ = 0$^{+}$ Isobaric Analogue State (IAS), fed by the Fermi transition, and a number of $T$ = 1, $J^{\pi}$ = 1$^{+}$ states due to Gamow-Teller transitions. In the present section we describe the procedure used for the determination of the Fermi and GT transition strengths, $B$(F) and $B$(GT), respectively. The peculiarities observed in the decay of each nucleus will be addressed in Section \ref{results}.

For each nucleus, for the proton and $\gamma$ peaks identified in the DSSSD and $\gamma$ spectra respectively, we determined the energies of the peaks (from the centroids obtained in the fits) and the intensities of the transitions (from the areas of the peaks). The $\beta$ feeding $I_\beta$ to the states populated in the daughter nucleus is then deduced.

In proton-rich nuclei the proton decay is expected to dominate for states well above ($>$1 MeV) the proton separation energy $S_p$ and so usually the $\beta$ feeding is readily inferred from the intensities of the proton peaks. However, the de-excitation of the $T$ = 2, IAS via proton decay is isospin-forbidden and it can only happen because of a $T$ = 1 isospin impurity in the IAS. In such cases competition between $\beta$-delayed proton emission and $\beta$-delayed $\gamma$ de-excitation from the IAS becomes possible, also because relatively low proton energies are usually involved. We indeed observe such competition in the decay of $^{48}$Fe, $^{52}$Ni and $^{56}$Zn. $^{56}$Zn is a special case where the competition becomes possible even at energies well above $S_p$. Thus for all three nuclei the intensities of the proton and $\gamma$ transitions from the IAS have to be added to get the correct $\beta$ feeding to the IAS, and hence the right amount of Fermi strength. Furthermore, in the case of $^{56}$Zn we observed the exotic $\beta$-delayed $\gamma$-proton decay in three cases \cite{PhysRevLett.112.222501}. Therefore for a proper determination of $B$(GT) of a given level the intensity deduced from the proton transition has to be corrected for the amount of indirect feeding coming from the $\gamma$ de-excitation.

The measured $T_{1/2}$ and $I_\beta$ were used to determine the $B$(F) and $B$(GT) values according to the equations:
\begin{equation}
  B(\mathrm{F}) = K~\frac{I_{\beta}^{IAS}(E_{IAS})}{f_{IAS}~T_{1/2}} \,,
  \label{Eq7}
\end{equation}

\begin{equation}
  B_j(\mathrm{GT}) = \frac{K}{\lambda^{2}}~\frac{I_{\beta}^{j}(E_j)}{f_j~T_{1/2}}\,,
  \label{Eq8}
\end{equation}
where the index $j$ indicates the daughter state at energy $E_j$, $f_j=f(Q_{\beta} - E_j,Z)$ is the Fermi factor \cite{Gove1971205}, \mbox{$K$ = 6143.6(17)} \cite{Fujita2011549} and $\lambda$ = -1.2701(25) \cite{Olive}.

\section{\label{results}Experimental results}

In this section we show the experimental results for the $T_z$ = -2 nuclei $^{48}$Fe, $^{52}$Ni and $^{56}$Zn. The total numbers of implantation events $N_{imp}$, the half-lives $T_{1/2}$ and the total proton-emission branching ratios $B_p$ are given in Table \ref{table}. 

\begin{table}[!h]
	\caption{Total numbers of implantation events, half-lives and total proton-emission branching ratios for the decays of $^{48}$Fe, $^{52}$Ni and $^{56}$Zn. In the lower half of the table results from Refs. \cite{Dossat200718,Faux1996167,PhysRevC.49.2440} are shown for comparison.}
	\label{table}
	\centering
	\begin{ruledtabular}
	  \begin{tabular}{c r r r c}
		  Isotope        & $N_{imp}$ & $T_{1/2}(ms)$ & $B_p$(\%) & Ref. \\ \hline
			$^{48}$Fe &    49763(268)  &        51(3)           &     14.4(7)       & this work\\
			$^{52}$Ni &   532054(729)  &      42.8(3)           &     31.1(5)       & this work\\
			$^{56}$Zn &      8861(94)  &      32.9(8)           &     88.5(26)      & this work\\ \hline
			$^{48}$Fe &    154241      &      45.3(6)           &     15.9(6)       & \cite{Dossat200718}\\
			                    &                      &        44(7)           &                         & \cite{Faux1996167}\\
			$^{52}$Ni &    272152      &      40.8(2)           &     31.4(15)      & \cite{Dossat200718}\\
			                    &                      &        38(5)           &                         & \cite{PhysRevC.49.2440}\\
			$^{56}$Zn &       630      &     30.0(17)           &     86.0(49)      & \cite{Dossat200718}\\ 
		  \end{tabular}
	\end{ruledtabular}
\end{table}

Our $T_{1/2}$ values are larger than those in the literature, nevertheless using the same analysis procedure we found a good agreement with Ref. \cite{Dossat200718} for the half-lives of other nuclei, such as $^{49}$Fe and $^{53}$Ni. In comparison with the previous study \cite{Dossat200718}, the present experiment achieved a higher energy resolution for protons, 70 keV FWHM, thanks to the use of a thinner DSSSD detector and better preamplifiers. The improved resolution, together with the increased statistics, allowed us to establish for the first time the decay scheme of $^{56}$Zn and to observe the exotic $\beta$-delayed $\gamma$-proton decay \cite{PhysRevLett.112.222501}.

New and detailed spectroscopic information has also been obtained on the $\beta$ decays of $^{48}$Fe and $^{52}$Ni because of the improved experimental conditions. The decay schemes of $^{48}$Fe and $^{52}$Ni have been enriched with many new states identified in the corresponding daughter nuclei, and the $\beta$-decay strengths have been extracted.

The details for each of the three nuclei under study are discussed in the following sections.

\subsection{\label{48Fe}Beta decay of $^{48}$Fe}

The $^{48}$Fe nucleus was first detected at GANIL \cite{Pougheon1987}. Results from decay studies were presented in Refs. \cite{Faux1996167,Dossat200718}.

\begin{figure}[!t]
  \centering
  \includegraphics[width=1.\columnwidth]{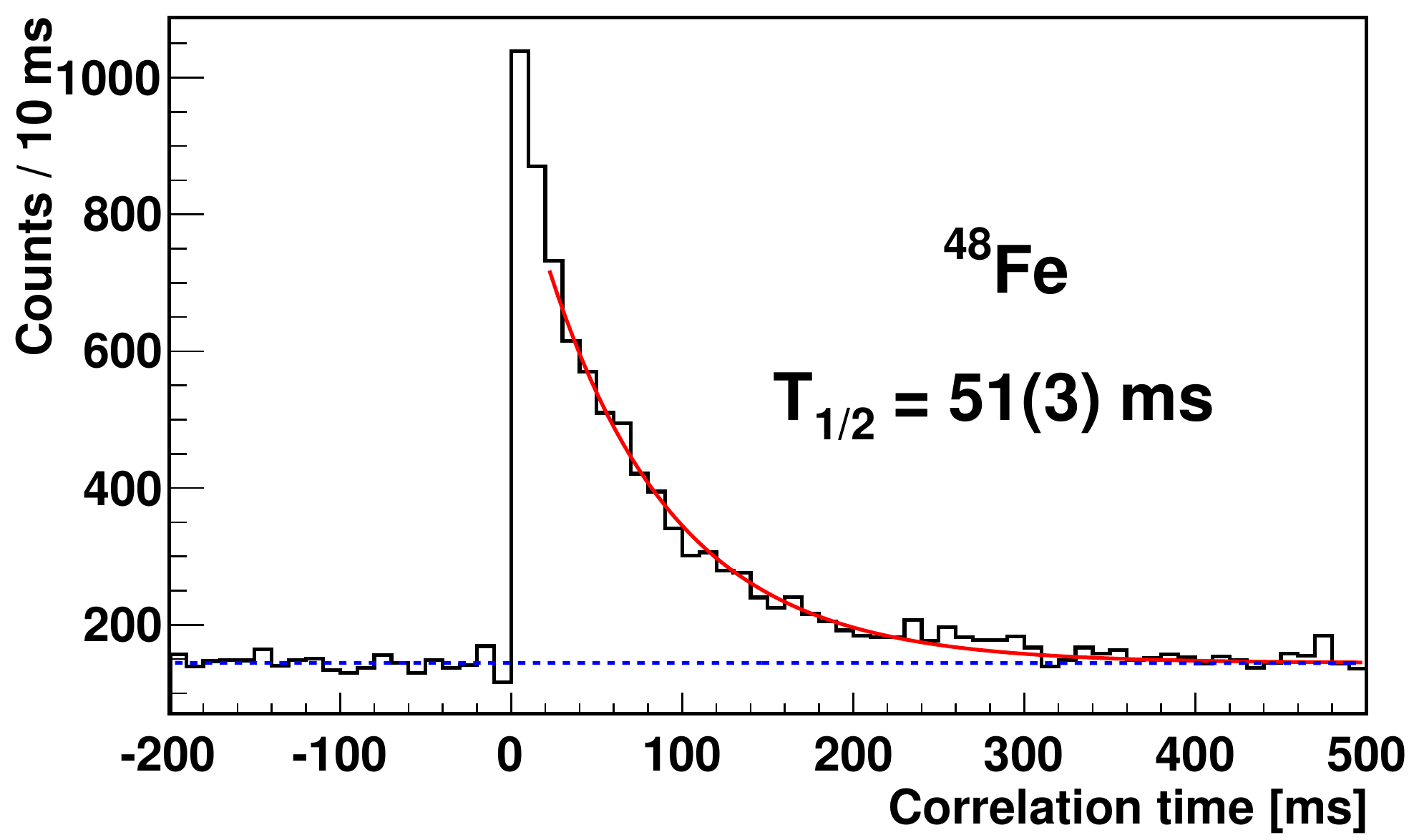}
  \vspace{-5.0 mm}
	\caption{Spectrum of the time correlations between all the $^{48}$Fe implants and each proton decay (DSSSD energy \mbox{$E_p>$ 0.94 MeV}, with the additional condition $E_p\notin$[1.85,2.23] MeV to remove the $^{49}$Fe contaminant).}
	\label{T48p}
  \vspace{-5.0 mm}
\end{figure}

Fig. \ref{T48p} shows the correlation-time spectrum for $^{48}$Fe obtained in our experiment, where proton decays with energy above 0.94 MeV have been selected in the DSSSD and the region corresponding to the $^{49}$Fe impurity (see below) has been removed. The half-life is determined by fitting the data including the $\beta$ decay of $^{48}$Fe (Eq. \ref{Eq1}) and the random correlation background (Section \ref{T}). A half-life $T_{1/2}$ = 51(3) ms is obtained. The maximum likelihood and least squares minimization methods gave the same result.

\begin{figure}[!h]
	\begin{minipage}{1.0\linewidth}
    \centering
    \includegraphics[width=1\columnwidth]{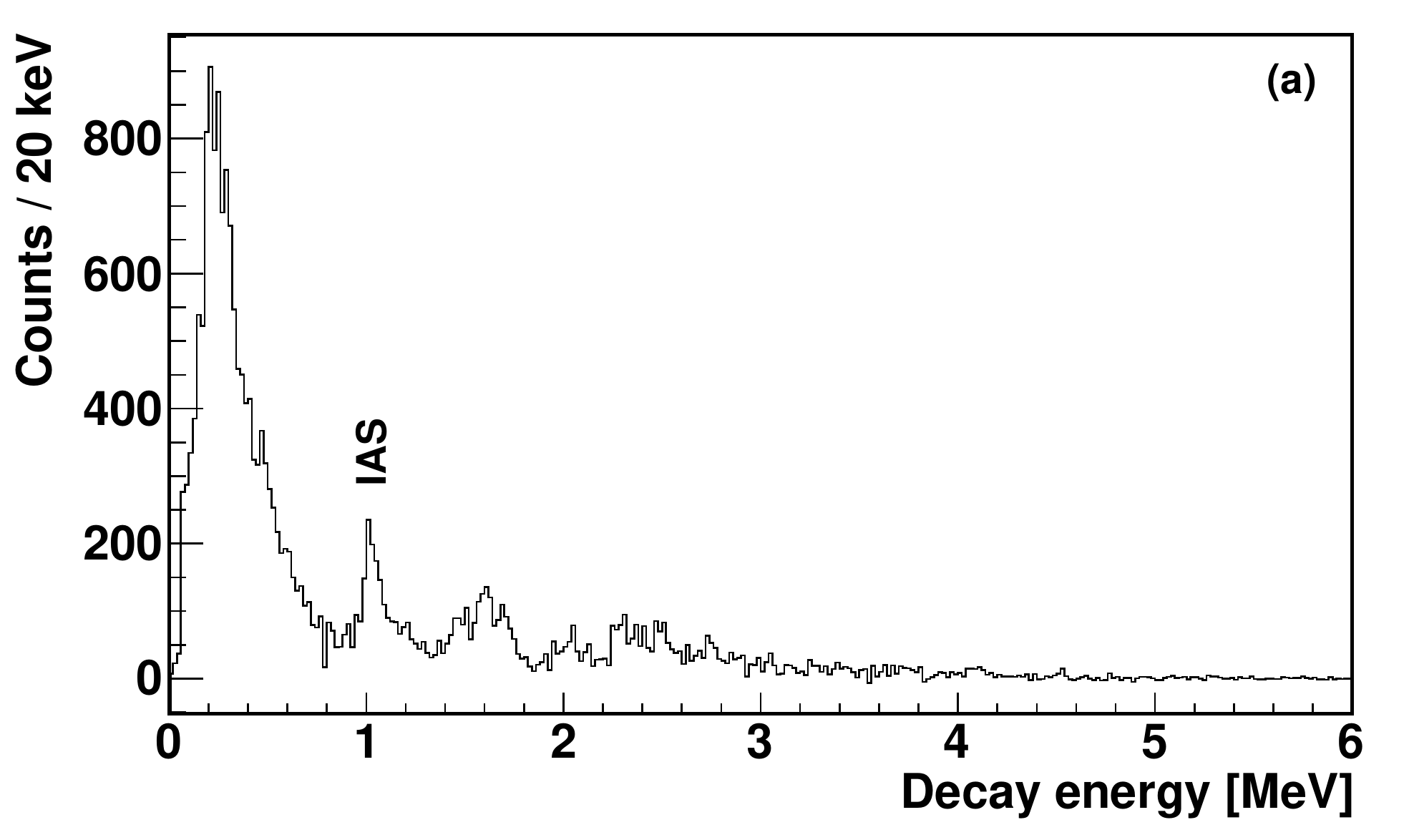}
	\end{minipage}
	\begin{minipage}{1.0\linewidth}
	  \centering
    \includegraphics[width=1\columnwidth]{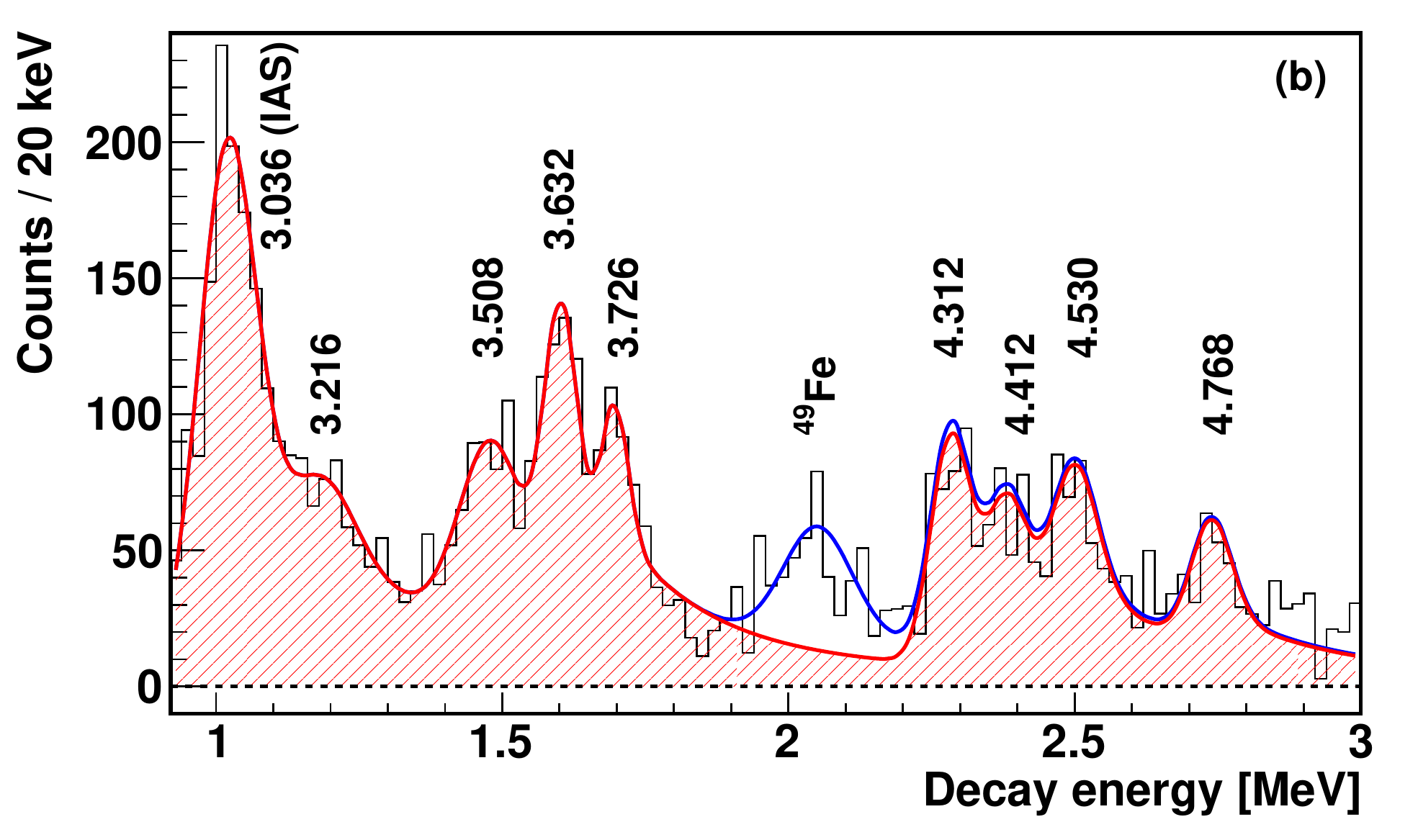}
	\end{minipage}
	\caption{(a) DSSSD charged-particle spectrum for decay events correlated with $^{48}$Fe implants.  (b) Fit of the peaks related to the proton emission following the $\beta$ decay of $^{48}$Fe. Peaks are labelled according to the corresponding excitation energies in the $\beta$-daughter $^{48}$Mn. The peak labelled as $^{49}$Fe (in blue) is due to a residual contamination from this isotope.}
	\label{dsssd48}
  \vspace{-5.0 mm}
\end{figure}

The charged-particle spectrum measured in the DSSSD for decay events correlated with $^{48}$Fe implants is shown in Fig. \ref{dsssd48}a. The large bump observed at low energy is due to the detection of $\beta$ particles not coincident with protons. Nine discrete peaks are identified above this bump and interpreted as being due to $\beta$-delayed proton emission. They are fitted as explained in Section \ref{DSSSD}. The fit is shown in Fig. \ref{dsssd48}b where the peaks are labelled according to the corresponding excitation energies in $^{48}$Mn, obtained as $E_X=E_p + S_p$ with $S_p$ = 2018(10) keV (see Section \ref{masses}). The proton decay of the IAS is identified at $E_X$ = 3.036(2) MeV as in Ref. \cite{Dossat200718}. 

The additional peak visible at $E_p\approx$ 2 MeV (in blue) is due to a residual contribution (3.0(3)\%) of the $^{49}$Fe contaminant. As explained in Ref. \cite{Dossat200718}, the background subtraction procedure cannot remove completely the activity of very strongly produced isotopes such as $^{49}$Fe. By imposing an increasingly restricted identification cut on $^{48}$Fe we can progressively get rid of the $^{49}$Fe contaminant, but at the price of poorer statistics. We have chosen the reasonable compromise of obtaining a good separation between the residual $^{49}$Fe peak and the $^{48}$Fe peaks. We have also checked that there is no visible contribution from $^{48}$Fe below the $^{49}$Fe peak. Hence in the construction of the correlation-time spectrum, besides imposing a DSSSD threshold of 0.94 MeV, we have also removed the energy region corresponding to the $^{49}$Fe peak by imposing $E_p\notin$[1.85,2.23] MeV. In this way we get a $B_p$ = 14.4(7)\%. Since in Ref. \cite{Dossat200718} the residual $^{49}$Fe impurity is not removed, they get a slightly different value, 15.9(6)\%. If we include the $^{49}$Fe peak we also get 15.1(7)\%.

\begin{figure}[!t]
  \centering
  \includegraphics[width=1\columnwidth]{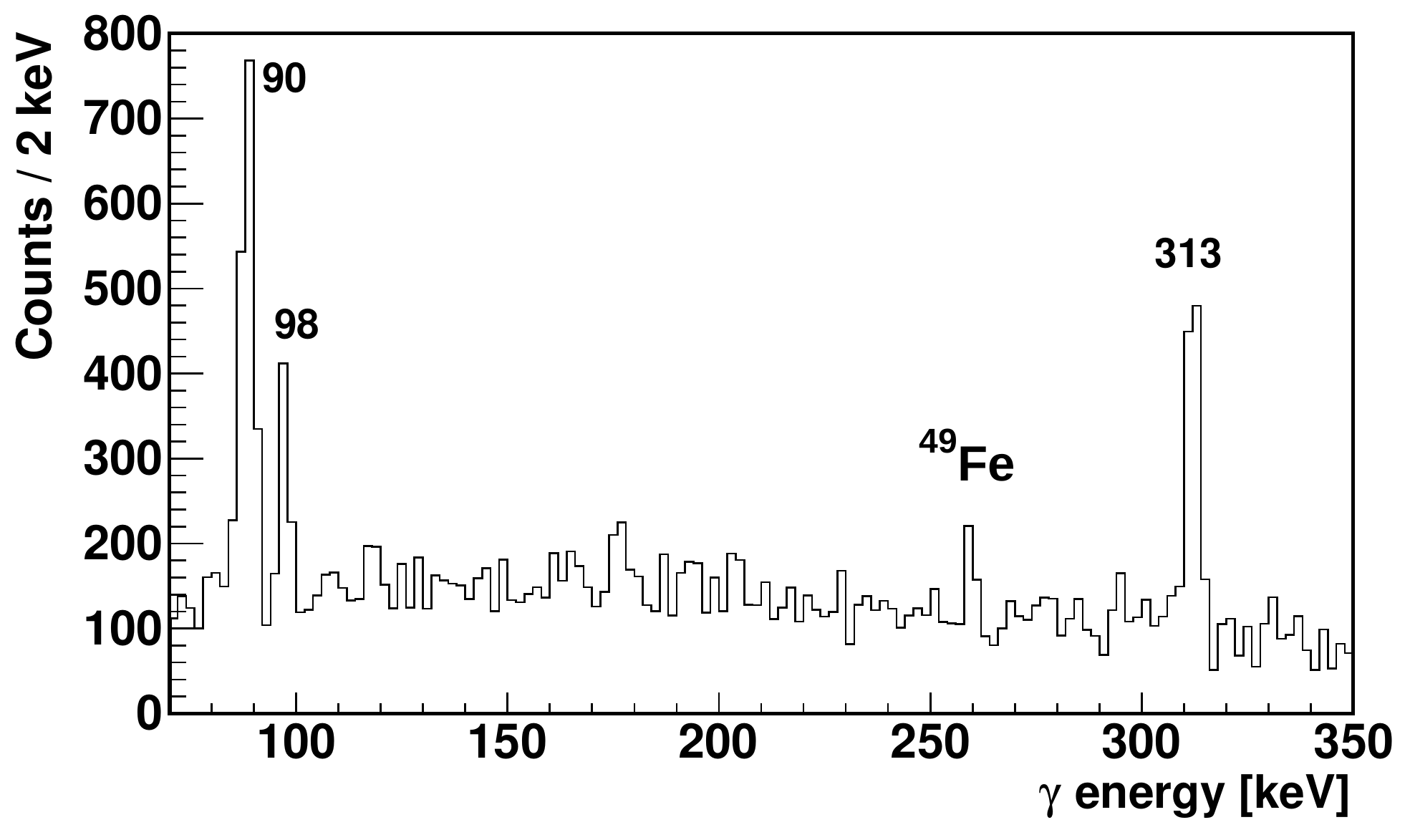}
  \vspace{-5.0 mm}
	\caption{$\gamma$-ray spectrum for decays correlated with $^{48}$Fe.}
	\label{gamma48}
\end{figure}

The $\gamma$-ray spectrum measured in coincidence with all the decays correlated with the $^{48}$Fe implants is shown in Fig. \ref{gamma48}. Three $\gamma$ lines are observed at 90, 98 and 313 keV. The fourth line visible in Fig. \ref{gamma48} at 261 keV comes from the residual $^{49}$Fe contaminant. A further $\gamma$ ray of 2634 keV is expected in the decay of $^{48}$Fe \cite{Dossat200718}, which we can only partially observe because of the problem affecting the low-amplification electronic chain (see Section \ref{gamma}). Thus for this $\gamma$ ray we took $I_{\gamma}$ = 30(5)\% from Ref. \cite{Dossat200718} to calculate the $\beta$-decay strength.

\begin{figure}[!t]
  \centering
  \includegraphics[width=1\columnwidth]{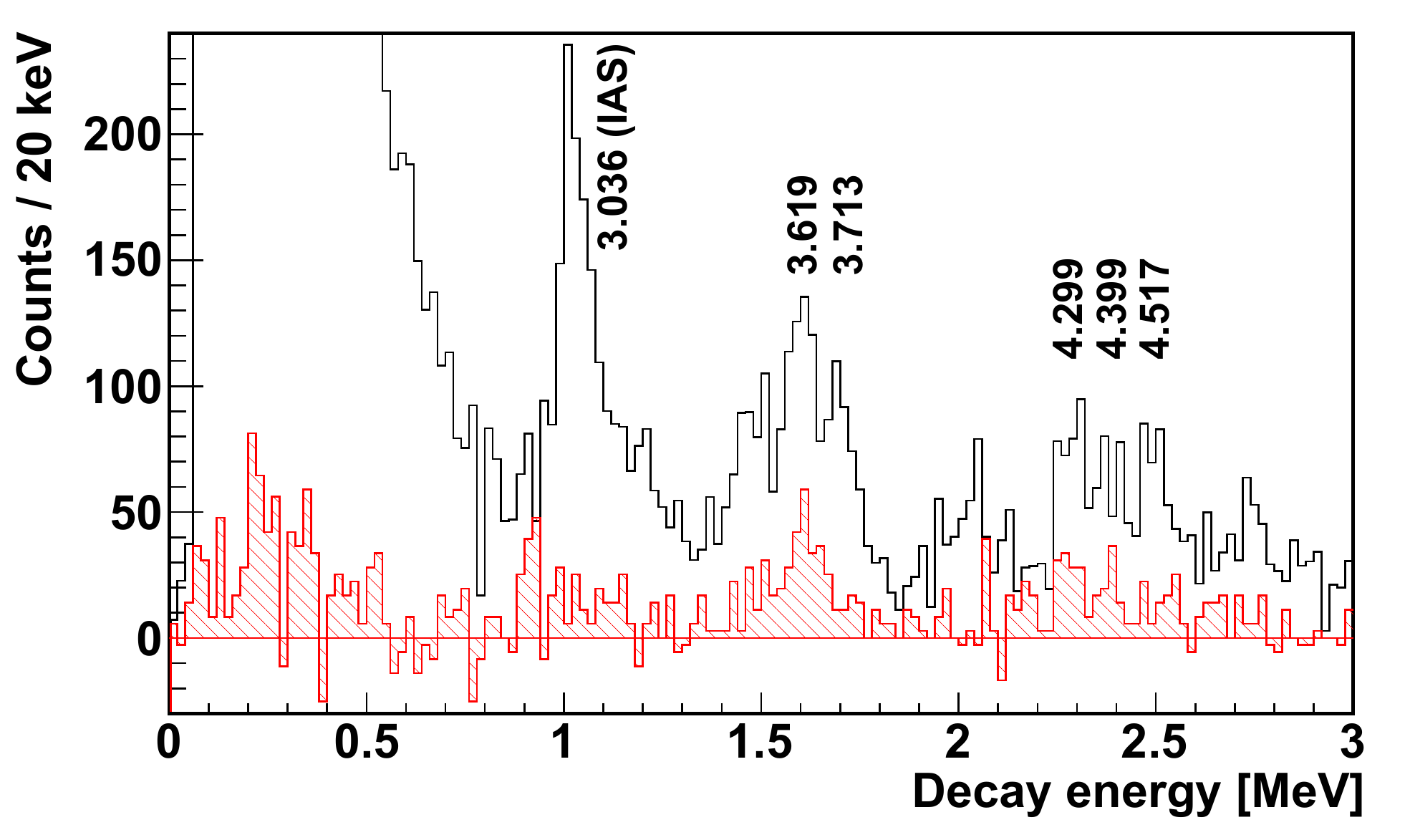}
  \vspace{-5.0 mm}
	\caption{DSSSD charged-particle spectrum correlated with $^{48}$Fe implants. The black-empty histogram is the full DSSSD spectrum. The dashed-red histogram is gated on the $\gamma$ line at 98 keV and scaled up to account for the $\gamma$ efficiency.}
	\label{gp98}
  \vspace{-5.0 mm}
\end{figure}

Gamma-proton coincidences were performed for all the $\gamma$ rays observed. As expected, the $\gamma$ lines at 90 and 313 keV are found to be coincident with the $\beta$ bump. The 98 keV line comes from the decay of the first excited state to the ground state in the proton-daughter $^{47}$Cr \cite{Dossat200718}.

\begin{table}[!t]
	\caption{Correction of the intensity of the $^{48}$Fe $\beta$-delayed proton emission due to the coincidences with the 98 keV $\gamma$ ray (see text). Centre-of-mass proton energies $E_p$, excitation energies $E_X$ in $^{48}$Mn, the intensities before correction $I_p^{i}$ (normalized to 100 decays), the corrections derived from the spectrum in coincidence with the 98 keV $\gamma$ ray, $\Delta I_p$(\%) (see Fig. \ref{gp98} and text), and the intensities after correction $I_p^{f}$.}
	\label{corrections}
	\centering
	\begin{ruledtabular}
	  \begin{tabular}{r r r l r}
 		  $E_p$(keV) & $E_X$(keV)   & $I_p^{i}$(\%) & $\Delta I_p$(\%) & $I_p^{f}$(\%) \\ \hline
			2499(10)      &    4517(14)     &      1.2(5)           &           +0.1           & 1.3(5)     \\
			2381(10)      &    4399(14)     &      0.8(4)           &           +0.2 -0.1   & 0.9(4)      \\
			2281(10)      &    4299(14)     &      1.3(3)           &           +0.1 -0.2   & 1.2(3)       \\  \hline
			1695(10)      &    3713(14)     &      0.7(2)           &           +0.6           & 1.3(2)       \\
			1601(10)      &    3619(14)     &      1.5(2)           & ~~~~~~~-0.6       & 0.9(3)        \\  \hline
			1018(10)      &  3036(2)$^b$ &      4.5(3)           &           +0.3           & 4.8(3)        \\
		  \end{tabular}
	\end{ruledtabular}
\raggedright{$^b$ IAS.}
\vspace{-5.0 mm}
\end{table}

It is worth noting that the pair of proton peaks at 3.619 and 3.713 MeV differs in energy by approximately 100 keV, close to the energy of the first excited state in $^{47}$Cr. The DSSSD spectrum gated on the $\gamma$ line at 98 keV, corrected by the corresponding $\gamma$ efficiency, is shown in Fig. \ref{gp98} (dashed-red histogram) and compared with the full DSSSD spectrum (black-empty histogram). The 98 keV $\gamma$ ray is found to be in coincidence with part of the 3.619 MeV proton peak, indicating that this part of the intensity (0.6\%), given in Table \ref{corrections} as $\Delta I_p$ in percent, comes from the decay of the observed 3.713 MeV level to the first excited state of $^{47}$Cr. Hence the intensities of the 3.619 and 3.713 MeV peaks have been corrected by shifting this 0.6\% intensity from the lower energy level to the upper one, as indicated in Table \ref{corrections}. At higher energy a triplet of peaks is observed (4.299, 4.399 and 4.517 MeV), the first two of which are also separated by around 100 keV, for which a small amount of intensity is also observed in coincidence with the 98 keV $\gamma$ ray (dashed-red histogram in Fig. \ref{gp98}). Thus the intensities of this triplet of peaks have been corrected in a similar way (see Table \ref{corrections}). Finally, in the $\gamma$-gated spectrum a small peak (of intensity 0.3\%) is seen at \mbox{$E_p\approx$ 0.9} MeV, which is attributed to the decay of the IAS to the first excited state in $^{47}$Cr, thus this intensity is added to the IAS one. Table \ref{corrections} shows the corrections and the intensities, before and after correcting.

\begin{figure}[!t]
  \centering
  \includegraphics[width=1\columnwidth]{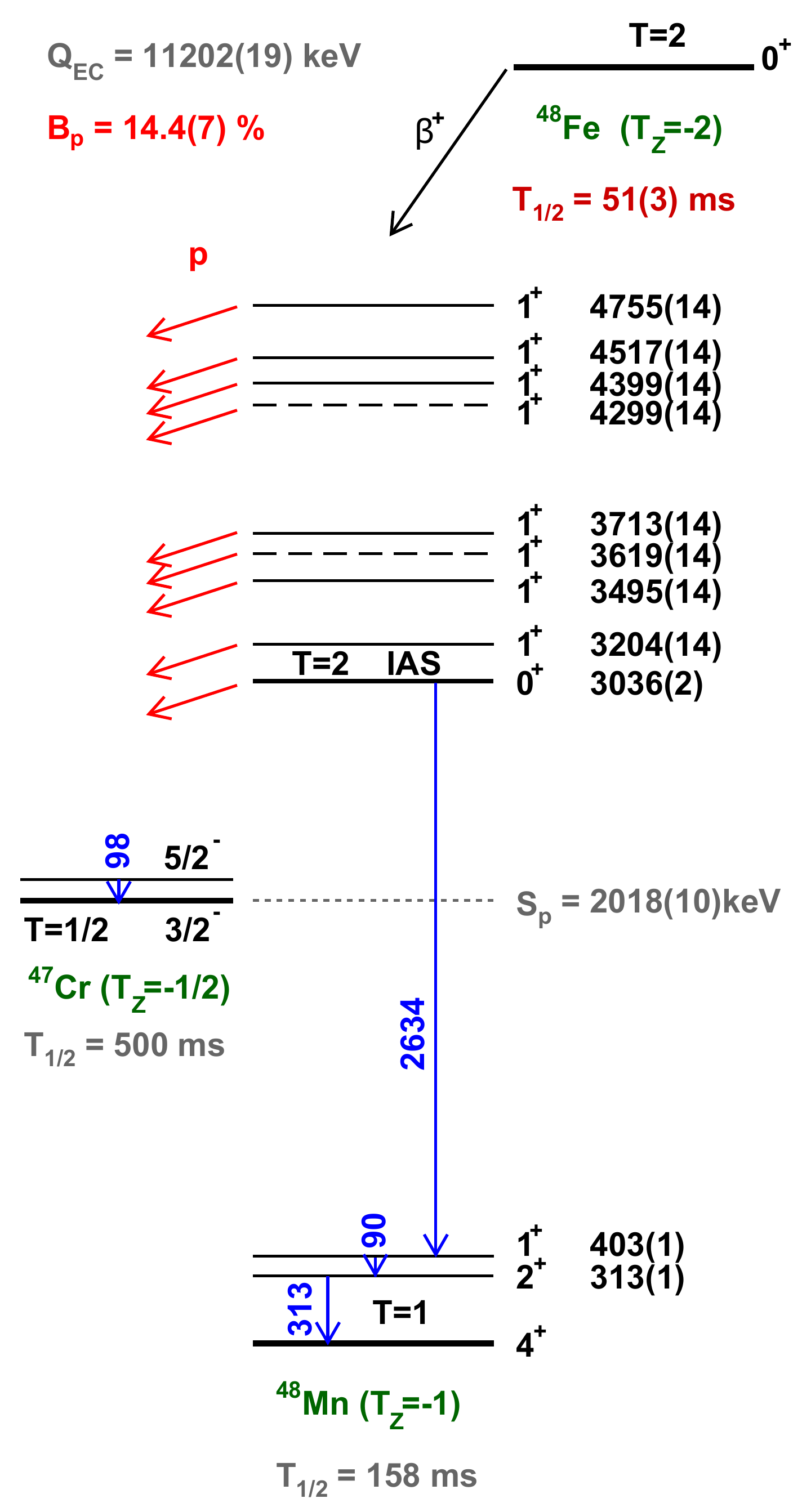}
  \vspace{-5.0 mm}
	\caption{Decay scheme of $^{48}$Fe deduced from the present experiment. Explanations on the dashed levels at 3.619 and 4.299 MeV are given in the text.}
	\label{decay48}
  \vspace{-5.0 mm}
\end{figure}

The weight of evidence outlined above supports the decay scheme shown in Fig. \ref{decay48}. There is an alternative explanation, namely that the levels at 3.619 and 4.299 MeV do not exist and the corresponding proton peaks are due to transitions from the 3.713 and 4.399 MeV levels, respectively, to the first excited state in $^{47}$Cr. To indicate this possibility we show the levels at 3.619 and 4.299 MeV with dashed lines in Fig. \ref{decay48}.

The coincidence with the 98 keV $\gamma$ ray selects proton-decay events only. The half-life associated with the $\gamma$ line at 98 keV agrees well with the $^{48}$Fe half-life obtained in Fig. \ref{T48p}. This is an additional confirmation that the fit in Fig. \ref{T48p} is not affected by some possible residual contribution from the $\beta$ tail in the region $E_p>$ 0.94 MeV.

Our observations are summarized in the $^{48}$Fe decay scheme in Fig. \ref{decay48} and in Table \ref{table48}, which gives the energies and intensities of the proton and $\gamma$ peaks, the $\beta$ feedings, the $B$(F) and $B$(GT) strengths (where we used $Q_{\beta}$ = 11202(19) keV determined in Section \ref{masses}). Fig. \ref{decay48} also shows the half-lives of the daughter nuclei \cite{tuli}.

The IAS decays by both proton emission and $\gamma$ de-excitation, via the $\gamma$ cascade at 2634, 90 and 313 keV. Indeed barrier-penetration calculations give a partial proton half-life of $t_{1/2}\sim10^{-17}$ s, two orders-of-magnitude smaller than the $\gamma$ de-excitation probability. Since the proton emission is isospin-forbidden, the $\gamma$ decay can compete with it. From our data, each 100 decays from the IAS divide into 14(2) proton decays to the ground state of $^{47}$Cr, and 86(19) $\gamma$ decays to the ground state of $^{48}$Mn. The two possible decay modes of the IAS give two independent determinations of the excitation energy of the IAS. From our measured proton energy and the experimental mass excess of the ground states in $^{47}$Cr and $^{48}$Mn \cite{Audi2012} we get $E_X^{IAS}$ = 3.066(170) MeV, while from the summing of the energies of the de-exciting $\gamma$ lines we obtain $E_X^{IAS}$ = 3.036(2) MeV. The two values agree well with each other and the second is a lot more precise.

\begin{table*}[!t]
	\caption{Summary of the results for the $\beta^{+}$ decay of $^{48}$Fe. Centre-of-mass proton energies $E_p$, $\gamma$-ray energies $E_{\gamma}$, and their intensities (normalized to 100 decays) $I_p$ and $I_{\gamma}$, respectively. Excitation energies $E_X$, $\beta$ feedings $I_{\beta}$, Fermi $B$(F) and Gamow-Teller $B$(GT) transition strengths to the $^{48}$Mn levels. The values for the 2634 keV $\gamma$ ray are taken from Ref. \cite{Dossat200718}.}
	\label{table48}
	\centering
	\begin{ruledtabular}
	  \begin{tabular}{r r r r r r r r}
			$E_p$(keV) & $I_p$(\%)& $E_{\gamma}$(keV)& $I_{\gamma}$(\%)& $E_X$(keV)& $I_{\beta}$(\%) & $B$(F)  & $B$(GT) \\ \hline
			2737(10) & 0.8(1)     &                            &                           &    4755(14) &       0.8(1)    &                & 0.10(2)   \\
			2499(10) & 1.3(5)     &                            &                           &    4517(14) &       1.3(5)    &                & 0.16(6)   \\
			2381(10) & 0.9(4)     &                            &                           &    4399(14) &       0.9(4)    &                & 0.10(4)   \\
			2281(10) & 1.2(3)     &                            &                           &    4299(14) &       1.2(3)    &                & 0.13(3)   \\
			1695(10) & 1.3(2)     &                            &                           &    3713(14) &       1.3(2)    &                & 0.10(2)   \\
			1601(10) & 0.9(3)     &                            &                           &    3619(14) &       0.9(3)    &                & 0.06(2)   \\
			1477(10) & 1.8(3)     &                            &                           &    3495(14) &       1.8(3)    &                & 0.12(2)   \\
			1186(10) & 1.0(3)     &                            &                           &    3204(14) &       1.0(3)    &                & 0.06(2)   \\
			1018(10) & 4.8(3)     &   2633.5(5)$^a$ &      30(5)$^a$  &    3036(2)$^b$ &     34.8(50) &    2.8(4)  &                  \\
											 &                &           90(1)        &       72(14)        &        403(1) &      42(15)    &                & 0.47(17)  \\
			                 &                &        313(1)         &       65(13)        &        313(1) &                     &                &                  \\
		\end{tabular}
	\end{ruledtabular}
	\raggedright{$^a$ From Ref. \cite{Dossat200718}. $^b$ IAS.}
  \vspace{-5.0 mm}
\end{table*}

The total $\beta$ feeding of the IAS is 34.8(50)\% and \mbox{$B$(F) = 2.8(4)}. We calculated that the $\beta$ feeding should be 49(3)\% to get the expected $B$(F) = $|N-Z|$ = 4. A possible explanation for the missing feeding could be that there are other weak $\gamma$ branches which are not observed. In addition, we took the intensity of the 2634 keV $\gamma$ line from Ref. \cite{Dossat200718} where there is also missing IAS feeding. 
 
Considering $B_p$ = 14.4(7) \% and the intensity of the 90 keV $\gamma$ line, 72(14) \%, we get a total $\beta$ feeding of 87(14)\%, thus globally there is a missing $\beta$ feeding of 13(14)\%. This value is compatible with the missing feeding in the IAS, however the sizable uncertainty lets us explore an additional hypothesis. In the mirror nucleus $^{48}$V two 1$^{+}$ states are observed at 2.288 and 2.406 MeV, also seen in a recent measurement of the charge exchange (CE) reaction $^{48}$Ti($^{3}$He,$t$)$^{48}$V \cite{Ela2015}. Due to the mirror symmetry, these states should also exist in $^{48}$Mn and they could explain part of the missing feeding. However we cannot make a conclusive statement about this hypothesis because we do not see any $\gamma$ line compatible with their population or de-excitation, although the statistics expected for these weak lines may place them below our sensitivity limit. In addition the corresponding proton decay from these 1$^{+}$ states is expected at $E_p$ around 300-400 keV and so the corresponding peaks, if they exist, would lie beneath the $\beta$ bump, making it impossible to identify them in the DSSSD spectrum. In Fig. \ref{gp98}, however, the coincidence with the 98 keV $\gamma$ ray suppresses the $\beta$ bump and a group of peaks is visible in the region 200-500 keV, compatible with the expected energies. We calculated the barrier-penetration half-life for the expected protons and found it to be in the range 10$^{-10}$-10$^{-7}$ s, while the partial half-life for $\gamma$ decay using the Weisskopf estimate is around 10$^{-15}$-10$^{-14}$ s. Therefore the $\gamma$ decay should dominate unless additional reasons, lying in the structure of the nuclear states involved, would favour the proton decay. 

\subsection{\label{52Ni}Beta decay of $^{52}$Ni}

$^{52}$Ni was observed for the first time at GANIL \cite{Pougheon1987}. The decay of $^{52}$Ni was studied in Refs. \cite{PhysRevC.49.2440,Dossat200718}.

\begin{figure}[!b]
  \centering
  \includegraphics[width=1.\columnwidth]{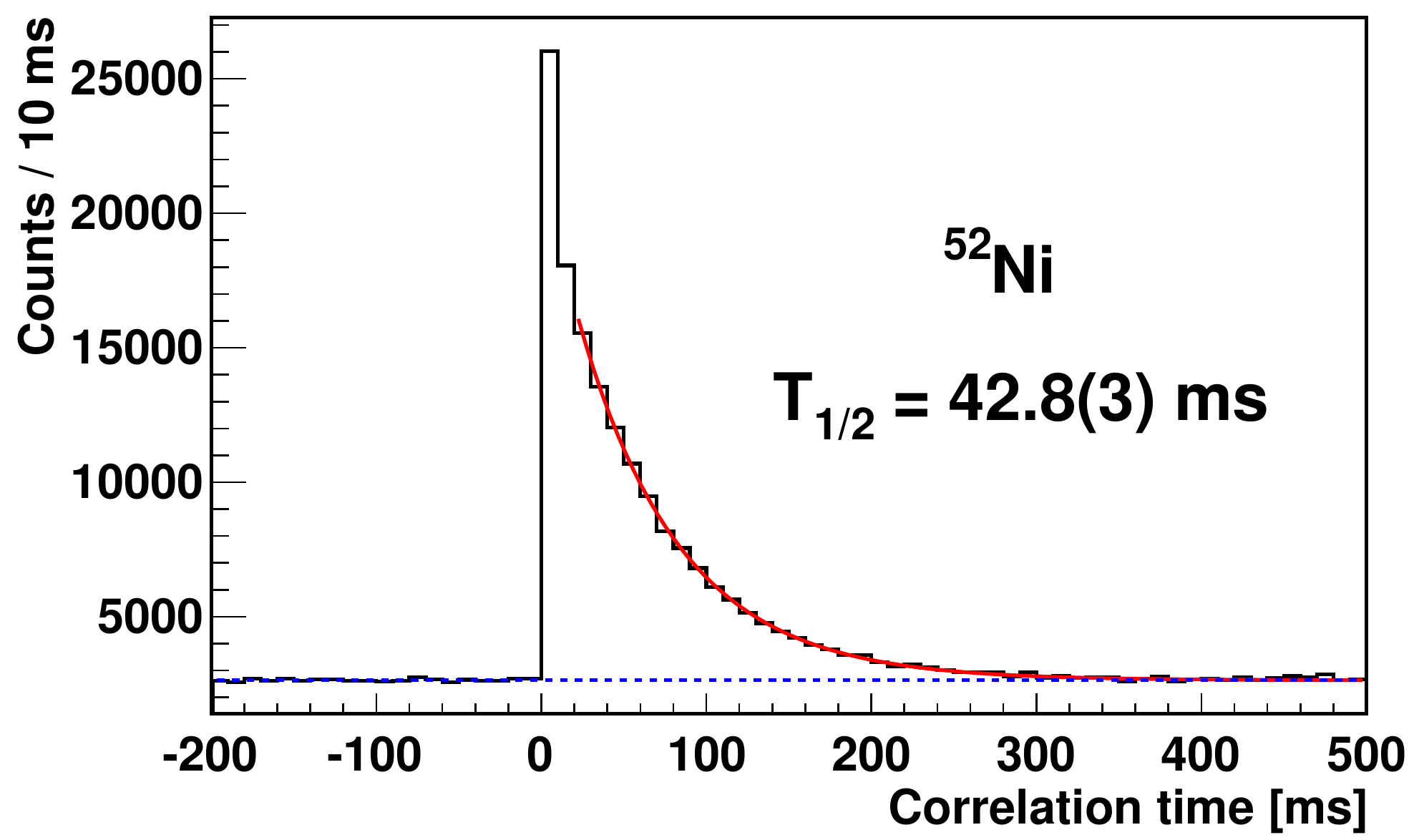}
  \vspace{-5.0 mm}
	\caption{Spectrum of the time correlations between each proton decay (DSSSD energy $>$ 0.98 MeV) and all $^{52}$Ni implants.}
	\label{T52p}
  \vspace{-5.0 mm}
\end{figure}

Fig. \ref{T52p} shows the correlation-time spectrum obtained for $^{52}$Ni in our experiment, where the proton decays have been selected by putting a threshold of 0.98 MeV on the energy in the DSSSD. The data are fitted with the function of Eq. \ref{Eq1}, including the $\beta$ decay of $^{52}$Ni and a constant background. A half-life of 42.8(3) ms is obtained. The maximum likelihood and least squares minimization methods gave the same result. The total proton branching ratio is determined as explained in Section \ref{Bp}. We obtain $B_p$ = 31.1(5)\%, in good agreement with the value 31.4(15)\% from Ref. \cite{Dossat200718}.

\begin{figure}[!b]
	\begin{minipage}{1.0\linewidth}
    \centering
    \includegraphics[width=1\columnwidth]{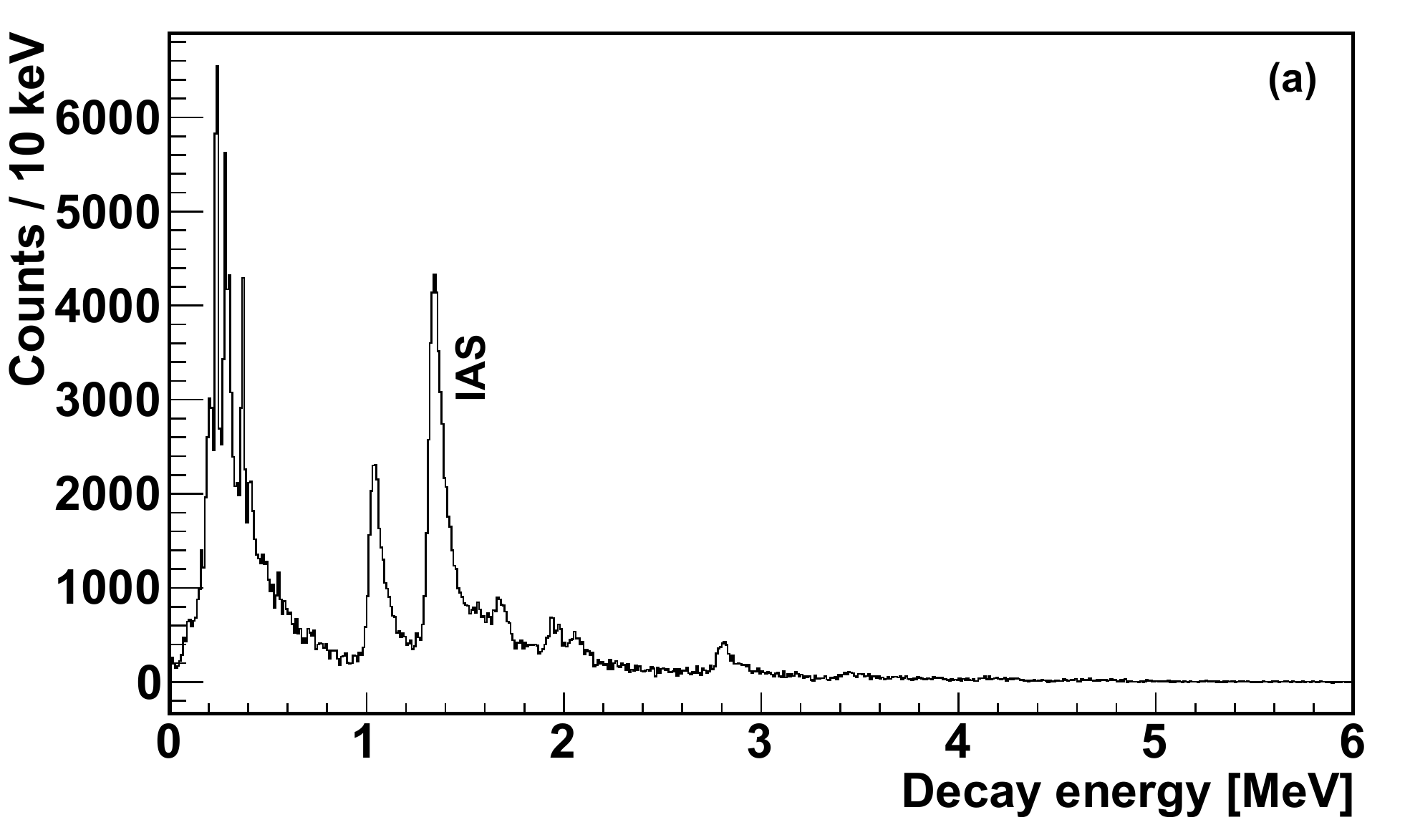}
	\end{minipage}
	\begin{minipage}{1.0\linewidth}
	  \centering
    \includegraphics[width=1\columnwidth]{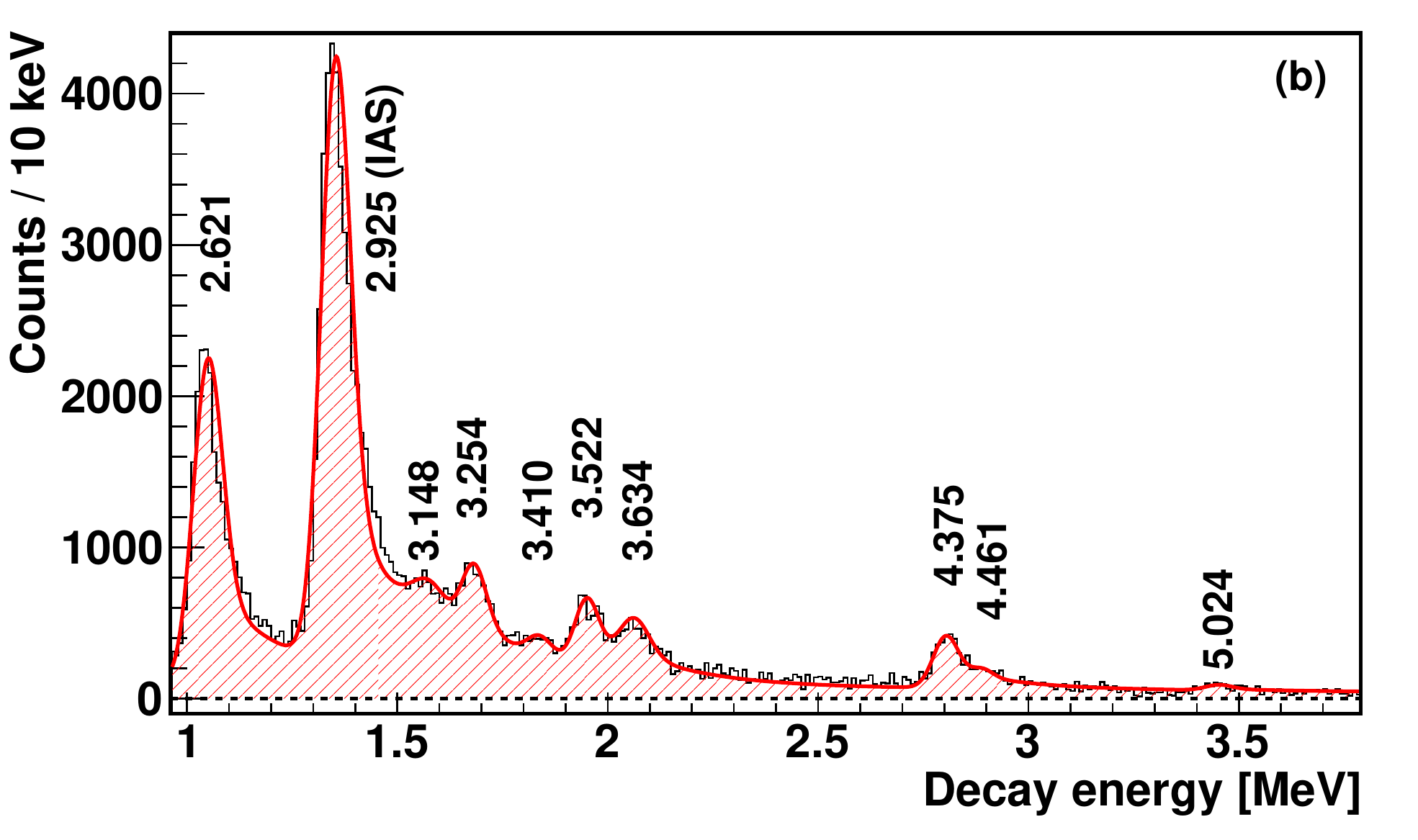}
	\end{minipage}
	\caption{(a) DSSSD charged-particle spectrum for decay events correlated with $^{52}$Ni implants.  (b) Fit of the peaks related to the proton emission following the $\beta$ decay of $^{52}$Ni. Peaks are labelled according to the corresponding excitation energies in the $\beta$-daughter $^{52}$Co.}
	\label{dsssd52}
  \vspace{-5.0 mm}
\end{figure}

Fig. \ref{dsssd52}a shows the DSSSD charged-particle spectrum obtained for decay events correlated with $^{52}$Ni implants. The bump observed below 1 MeV is attributed to $\beta$ particles not coincident with protons (the structure in the bump comes from the different thresholds of the strips in the DSSSD). Ten discrete peaks are identified above this bump and interpreted as being due to $\beta$-delayed proton emission. The fit to these peaks, performed as in Section \ref{DSSSD}, is shown in Fig. \ref{dsssd52}b. The peaks are labelled according to the corresponding excitation energies in $^{52}$Co, obtained by adding \mbox{$S_p$ = 1574(51) keV} (see Section \ref{masses}) to the measured proton energy $E_p$ in each case.

The DSSSD spectrum can be compared to the spectrum obtained from the mirror CE experiment, the reaction $^{52}$Cr($^{3}$He,$t$)$^{52}$Mn (see Fig. 45a in Ref. \cite{Fujita2011549}). All the dominant transitions are observed in both spectra, showing a good isospin symmetry. In detail, the $^{52}$Mn peaks seen in the CE spectrum at $E_X$ = 2.636, 3.585, 4.390 and 5.090 MeV correspond to the $^{52}$Co peaks seen in the DSSSD spectrum at $E_X$ = 2.622, 3.523, 4.376 and 5.024 MeV. In addition we see other small peaks at $E_X$ = 3.148, 3.254, 3.410, 3.634 and 4.462 MeV which seem to be only very weakly populated in the mirror CE process.

Moreover, we see a strong peak at $E_X$ = 2.926 MeV which is identified as the proton decay of the $^{52}$Co IAS, as in Ref. \cite{Dossat200718}. The higher energy resolution of the CE reaction allows the separation of two $^{52}$Mn states lying close in energy, at 2.875(1$^+$) and 2.938(0$^+$) MeV \cite{Fujita2011549}. Hence the peak seen in $^{52}$Co at 2.926 MeV could also contain a contribution from an unresolved 1$^+$ level, which is expected to be small since the 0$^+$ contribution is enhanced in the $\beta$ decay in comparison to CE \cite{Fujita2011549}. We neglected this possible contribution in the calculation of $B$(F).

\begin{figure}[!t]
	\begin{minipage}{1.0\linewidth}
    \centering
    \includegraphics[width=1\columnwidth]{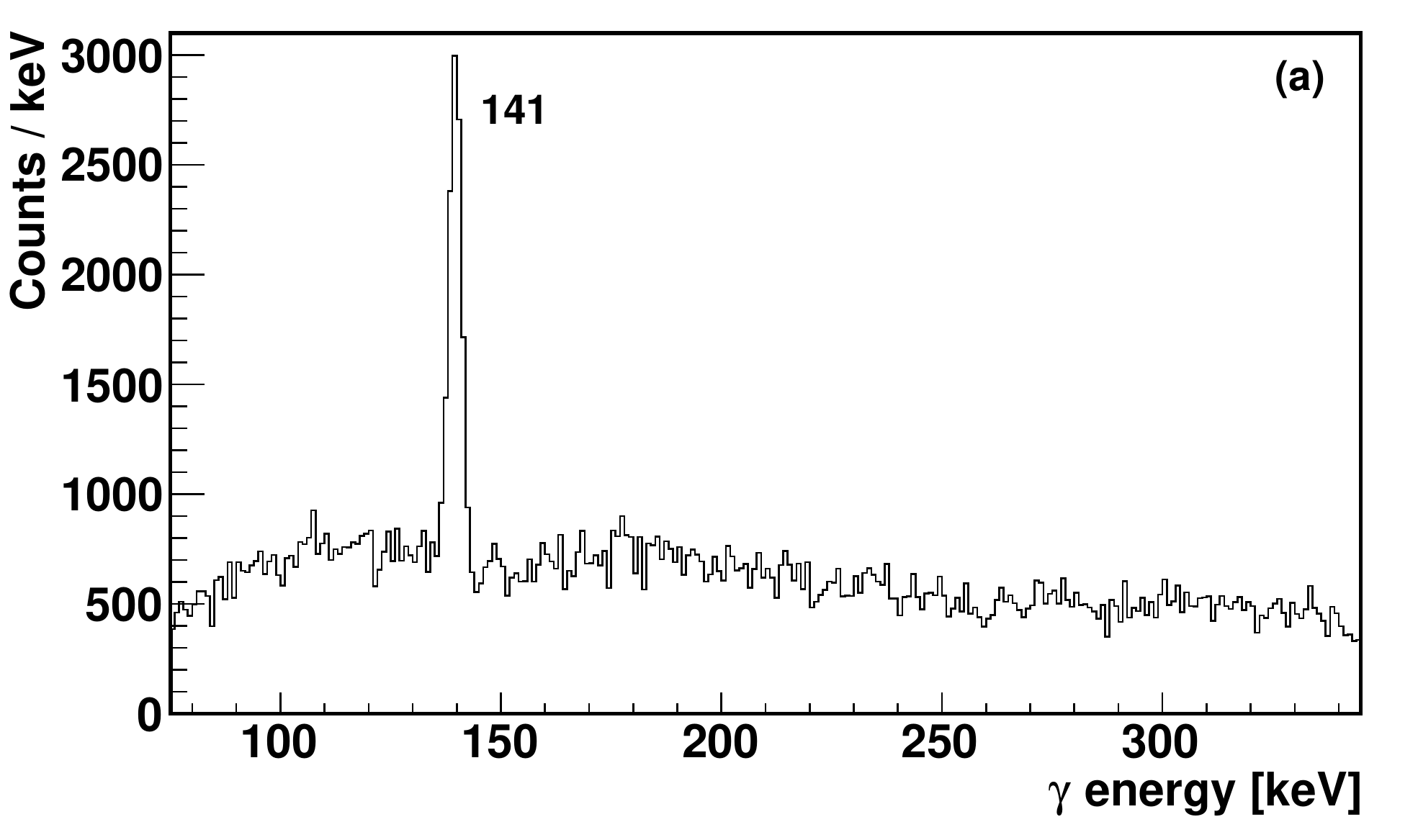}
	\end{minipage}
	\begin{minipage}{1.0\linewidth}
	  \centering
    \includegraphics[width=1\columnwidth]{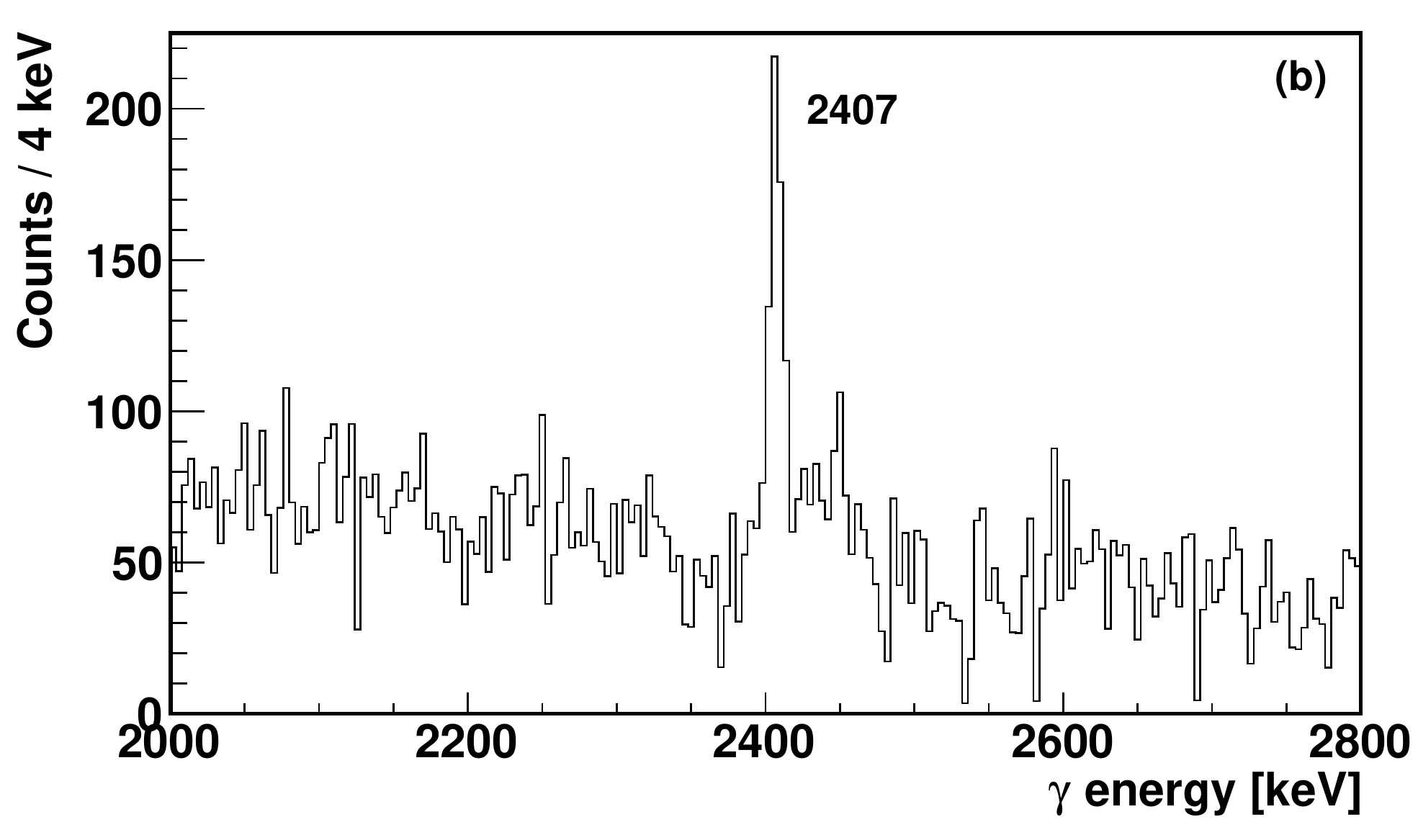}
	\end{minipage}
	\caption{$\gamma$-ray spectrum for decay events correlated with $^{52}$Ni implants. (a) High-amplification spectrum showing the $\gamma$ line at 141 keV. (b) Low-amplification spectrum showing the $\gamma$ line at 2407 keV.}
	\label{gamma52}
  \vspace{-5.0 mm}
\end{figure}

The $\gamma$-ray spectrum observed for the decay of $^{52}$Ni is shown in Fig. \ref{gamma52} (where the high-amplification and low-amplification spectra are shown in the panels a and b, respectively). Two $\gamma$ lines are observed at 141 and 2407 keV, also seen in Ref. \cite{Dossat200718} at 142 and 2418 keV, respectively. We determine for the first time the intensity of the 141 keV line, $I_{\gamma}$ = 43(8)\%. This line is found to be coincident with the $\beta$ bump, as expected. Even if the shape of the 2407 keV peak is distorted, due to the problem affecting the low-amplification electronic chain (see Section \ref{gamma}), the high statistics allowed us to extract an intensity $I_{\gamma}$ = 42(10)\% which agrees well with the value 38(5)\% from Ref. \cite{Dossat200718}.

\begin{figure}[!t]
  \centering
  \includegraphics[width=1\columnwidth]{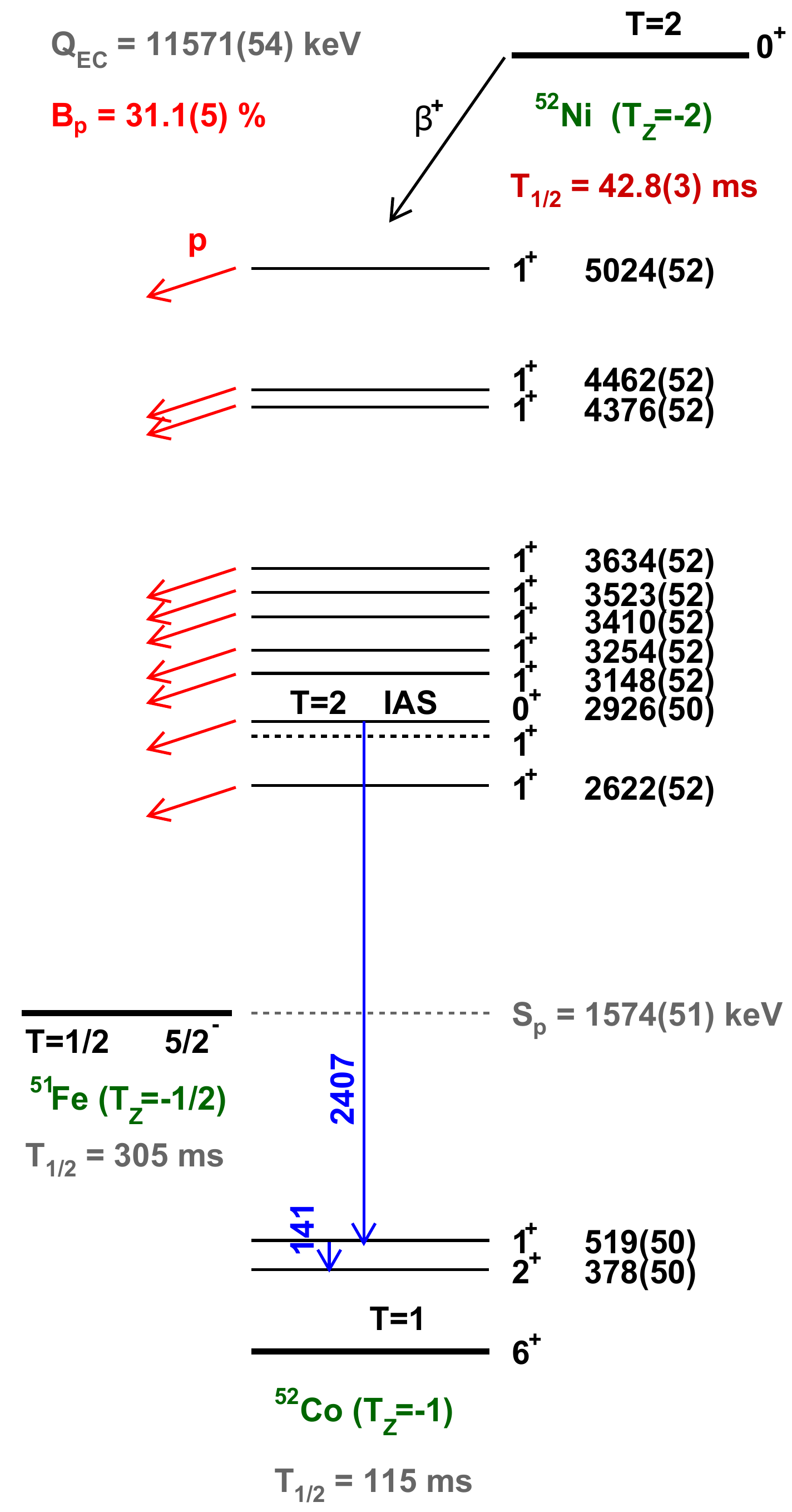}
  \vspace{-5.0 mm}
	\caption{Decay scheme of $^{52}$Ni deduced from the present experiment. The energy of the 2$^+$ level is assumed from the mirror $^{52}$Mn. The dashed 1$^{+}$ state has been observed in $^{52}$Mn.}
	\label{decay52}
  \vspace{-5.0 mm}
\end{figure}

Our observations are summarized in the $^{52}$Ni decay scheme shown in Fig. \ref{decay52} and in Table \ref{table52}, which gives the energies and intensities of the proton and $\gamma$ peaks, the $\beta$ feedings, the $B$(F) and $B$(GT) strengths (for which we used $Q_{\beta}$ = 11571(54) keV determined in Section \ref{masses}). Fig. \ref{decay52} also shows the half-lives of the daughter nuclei \cite{tuli}. The total $\beta$ feeding of the IAS is 56(10)\% and \mbox{$B$(F) = 4.1(8)}, consistent with the expected \mbox{$B$(F) = 4}. The 2.622 MeV level takes most of the GT strength. From $B_p$ and the intensity of the 141 keV $\gamma$ line we get a total $\beta$ feeding of 74(8)\%, thus globally there is a missing $\beta$ feeding of 26(8)\% which could belong to other unobserved weak $\gamma$ or proton branches.

\begin{table*}[!t]
	\caption{Summary of the results for the $\beta^{+}$ decay of $^{52}$Ni. Centre-of-mass proton energies $E_p$, $\gamma$-ray energies $E_{\gamma}$, and their intensities (normalized to 100 decays) $I_p$ and $I_{\gamma}$, respectively. Excitation energies $E_X$, $\beta$ feedings $I_{\beta}$, Fermi $B$(F) and Gamow-Teller $B$(GT) transition strengths to the $^{52}$Co levels.}	
	\label{table52}
	\centering
	\begin{ruledtabular}
	  \begin{tabular}{r r r r r r r r}
			 $E_p$(keV) & $I_p$(\%)& $E_{\gamma}$(keV)& $I_{\gamma}$(\%)& $E_X$(keV)& $I_{\beta}$(\%) & $B$(F)  & $B$(GT) \\ \hline
			3451(10) & 0.11(1)     &                            &                           &    5024(52) &       0.11(1)    &                & 0.017(2)   \\
			2888(10) & 0.18(2)     &                            &                           &    4462(52) &       0.18(2)    &                & 0.020(3)   \\
			2802(10) & 1.01(3)     &                            &                           &    4376(52) &       1.01(3)    &                & 0.106(6)   \\
			2061(10) & 1.14(3)     &                            &                           &    3634(52) &       1.14(3)    &                & 0.078(5)   \\
			1949(10) & 1.28(3)     &                            &                           &    3523(52) &       1.28(3)    &                & 0.082(5)   \\
			1836(10) & 0.42(3)     &                            &                           &    3410(52) &       0.42(3)    &                & 0.025(2)   \\
			1681(10) & 1.50(4)     &                            &                           &    3254(52) &       1.50(4)    &                & 0.082(5)   \\
			1575(10) & 1.17(4)     &                            &                           &    3148(52) &       1.17(4)    &                & 0.060(4)   \\
			1352(10) & 13.7(2)     &      2407(1)        &         42(10)       &    2926(50)$^a$ &   56(10)  &  4.1(8)   &                 \\
			1048(10) & 7.30(9)     &                            &                           &    2622(52) &       7.30(9)    &                & 0.28(1)   \\
											 &                  &        141(1)         &         43(8)         &      519(50) &          1(13)    &                & 0.01(15)  \\
		\end{tabular}
	\end{ruledtabular}
	\raggedright{$^a$ IAS}
  \vspace{-5.0 mm}
\end{table*}

The IAS in $^{52}$Co decays 75(23)\% of the time by $\gamma$ rays and 25(5)\% by (isospin-forbidden) proton emission to the ground state in $^{51}$Fe ($J^{\pi}$ = 5/2$^{-}$). The population of the first excited state at 253 keV (7/2$^{-}$) in $^{51}$Fe via proton decay would be followed by the emission of a $\gamma$ ray at 253 keV that we do not see. The $\gamma$ branch from the IAS proceeds via the $\gamma$ rays of 2407 and 141 keV which populate the levels in $^{52}$Co at 519 and 378 keV, respectively, while no $\gamma$ rays are observed from the 378 keV (2$^+$) level. As explained in detail in Section \ref{masses}, we assume for this level an energy of 378(50) keV from the value in the mirror nucleus $^{52}$Mn, 377.749(5) keV \cite{PhysRevC.7.677}, fixing the excitation energies for the $^{52}$Co levels. Barrier-penetration calculations give a partial proton half-life of $t_{1/2}\sim10^{-15}$ s for the decay of the IAS to the ground state in $^{51}$Fe, and the calculated Weisskopf transition probability for the $\gamma$ decay of the IAS to the 1$^{+}$ level at 519 keV is of the same order-of-magnitude. In the mirror $^{52}$Mn the 2$^+$ level at 378 keV is an isomer with a half-life of 21.1 min. Hence the 2$^+$ level in $^{52}$Co is also likely to be an isomeric state decaying by $\beta$ emission to $^{52}$Fe.

\subsection{\label{56Zn}Beta decay of $^{56}$Zn}

The $^{56}$Zn nucleus was observed for the first time in 1999 in an experiment carried out at GANIL \cite{Giovinazzo2001}. However, before the present experiment there was only a little information on the decay of $^{56}$Zn and the excited states of its daughter $^{56}$Cu. The observation of $\beta$-delayed proton emission was reported in Ref. \cite{Dossat200718}, but only by improving the statistics and energy resolution has it been possible to perform a fine study of the energy levels in $^{56}$Cu. Moreover, $\beta$-delayed $\gamma$ rays were reported for the first time in the present experiment \cite{PhysRevLett.112.222501}.

The correlation-time spectrum for $^{56}$Zn with selection of the proton emission (DSSSD energy above 0.8 MeV) is shown in Fig. \ref{T56p}. It has already been discussed in Section \ref{Tfit}. A least squares fit to the data using the function in Eq. \ref{Eq1} gives a half-life of $T_{1/2}$ = 32.9(8) ms.

Fig. \ref{dsssdbg}c shows the DSSSD charged-particle spectrum obtained for decay events correlated with $^{56}$Zn implants. In this case only a small number of $\beta$ particles not coincident with protons are observed below 0.8 MeV, while above this energy the decay is dominated by $\beta$-delayed proton emission. The fit of the six proton peaks identified, shown in Fig. \ref{MC}b, was performed as explained in Section \ref{DSSSD}. In the figure the peaks are labelled according to the corresponding excitation energies in $^{56}$Cu, obtained by adding $S_p$ = 560$^\#$(140$^\#$) keV \cite{Audi2003} (see Section \ref{masses}) to the measured proton energy $E_p$.

The total proton branching ratio, determined according to the procedure of Section \ref{Bp}, is $B_p$ = 88.5(26)\%, in good agreement with the value 86.0(49)\% reported in Ref. \cite{Dossat200718}. Thanks to the higher energy resolution, we were able to identify the first proton peak at $E_p$ = 0.831 MeV ($E_X$ = 1.391 MeV), while in Ref. \cite{Dossat200718} it was assumed proton emission only occurred above 0.9 MeV.

The comparison of the DSSSD spectrum with the mirror spectrum obtained in the $^{56}$Fe($^{3}$He,$t$)$^{56}$Co CE reaction \cite{PhysRevC.88.054329} has already been discussed in Ref. \cite{PhysRevLett.112.222501}. There is a remarkable isospin symmetry. The $^{56}$Cu levels seen in the DSSSD spectrum at $E_X$ = 1.691, 2.537, 2.661 and 3.508 MeV correspond to the $^{56}$Co levels seen in the CE spectrum at 1.720, 2.633, 2.729 and 3.599 MeV. The broader $^{56}$Cu peak at 3.423 MeV contains, unresolved \cite{PhysRevLett.112.222501}, at least two of the three states seen in $^{56}$Co at 3.432, 3.496 and 3.527 MeV. In addition, the $^{56}$Cu level at 1.391 MeV is the 0$^+$ antianalogue state \cite{Kampp1978} and corresponds to the $^{56}$Co level at 1.451 MeV, which is not populated in the CE reaction. The 2.633 MeV level is also very-weakly populated in the CE experiment. Thus we expect that the 1.391 and 2.537 MeV levels will only receive a small amount of feeding in the $\beta$ decay and are populated only indirectly by $\gamma$ decay from the levels above. This has been taken into account in the calculation of the $\gamma$ de-excitation of the IAS \cite{PhysRevLett.112.222501}.

\begin{figure}[!t]
  \centering
  \includegraphics[width=1\columnwidth]{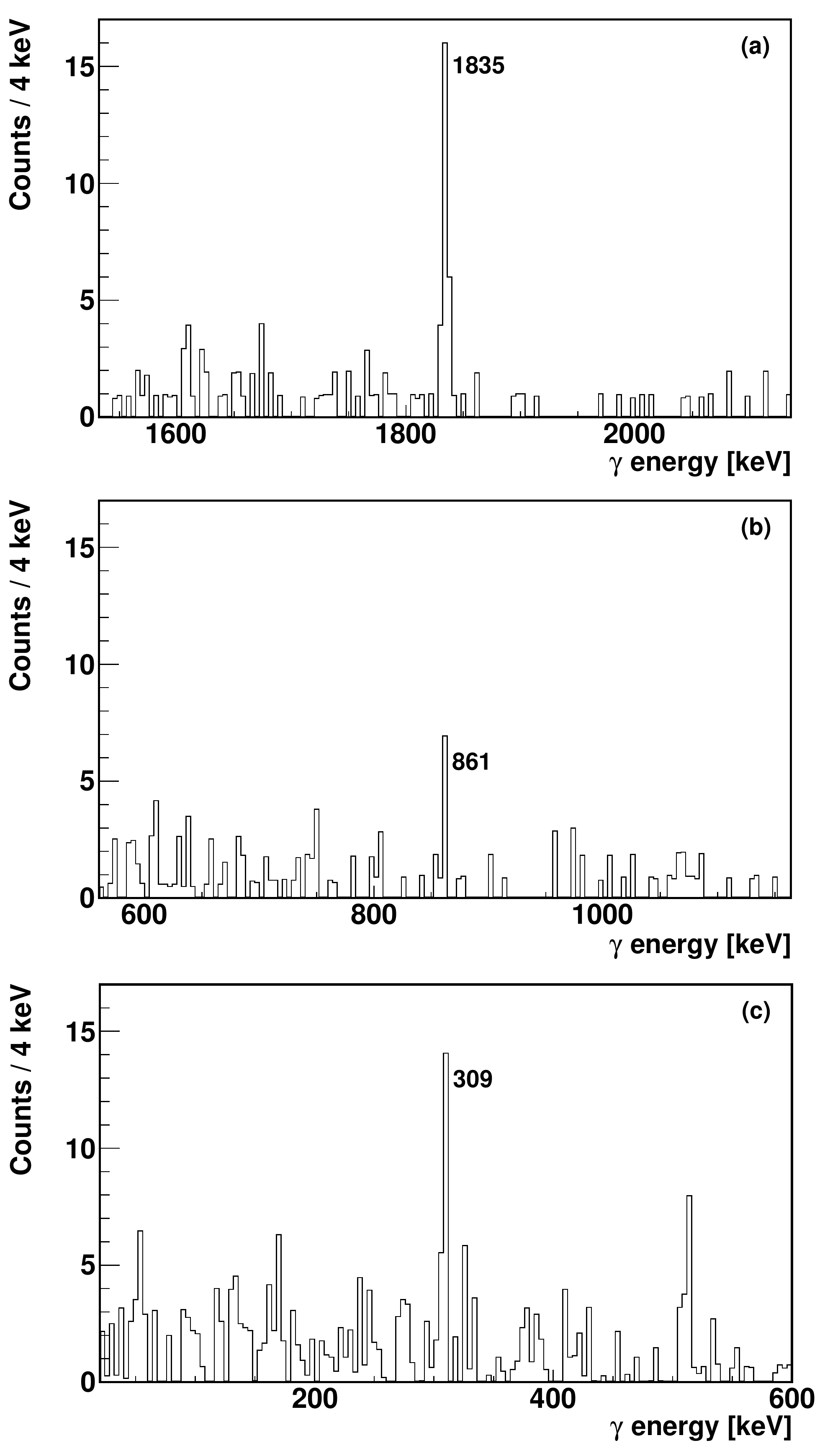}
  \vspace{-5.0 mm}
	\caption{$\gamma$-ray spectrum for decay events correlated with $^{56}$Zn implants in coincidence with protons from the levels at (a) 1.691, (b) 2.661 and (c) 1.391 MeV.}
	\label{gamma56}
  \vspace{-5.0 mm}
\end{figure}

A $\gamma$-ray spectrum for the decay of $^{56}$Zn has been measured for the first time. In the spectrum, shown in Fig. \ref{gammabg}c, a $\gamma$ ray at 1835 keV is observed. Two additional $\gamma$ rays have been identified at 309 and 861 keV from $\gamma$-proton coincidences. Fig. \ref{gamma56} shows the three $\gamma$ lines.

\begin{figure}[!t]
  \centering
  \includegraphics[width=1\columnwidth]{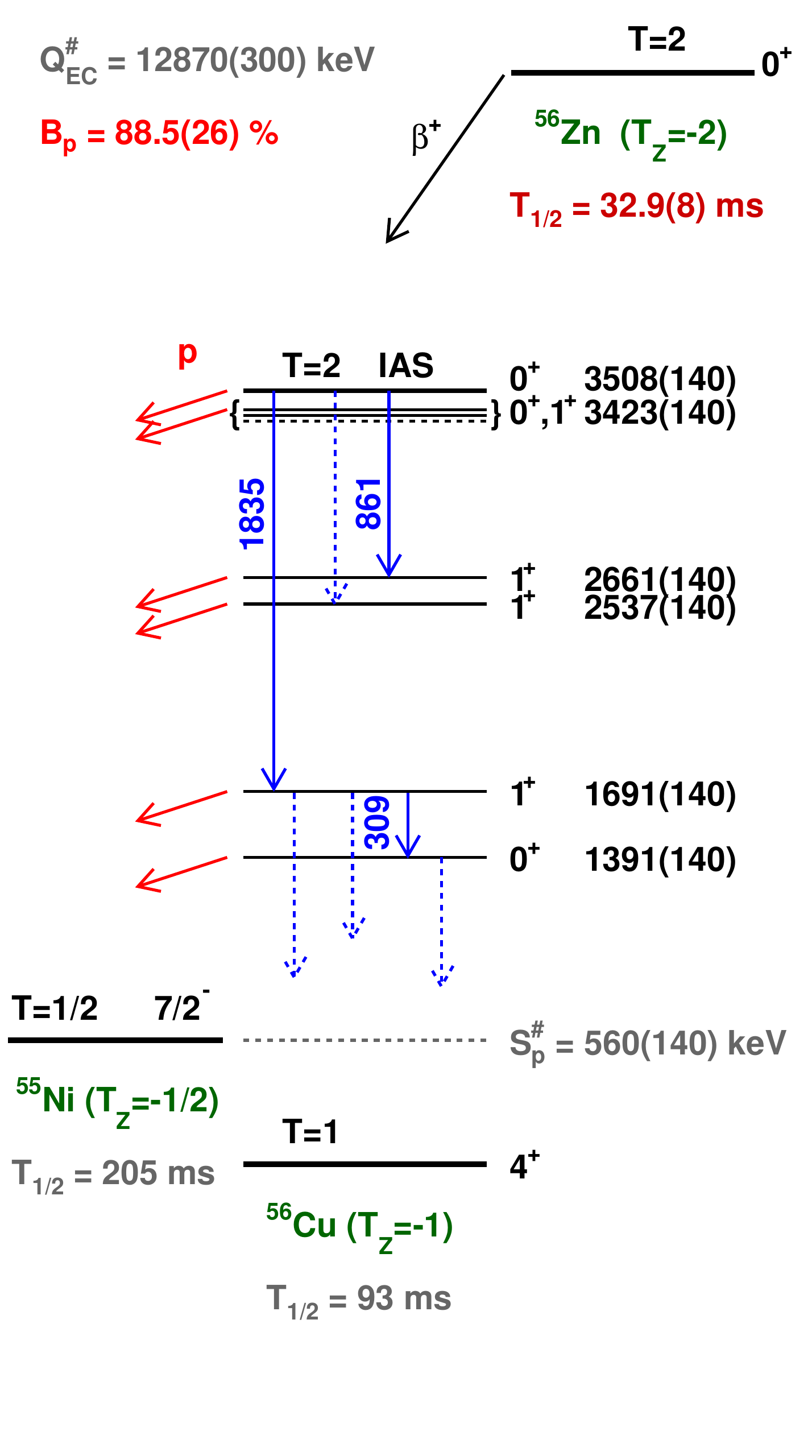}
  \vspace{-18.0 mm}
	\caption{Decay scheme of $^{56}$Zn deduced from the results of the present experiment. The dashed lines represent transitions corresponding to those observed in the mirror $^{56}$Co nucleus.}
	\label{decay56}
  \vspace{-5.0 mm}
\end{figure}

The $^{56}$Zn decay scheme is shown in Fig. \ref{decay56}, which also shows the half-lives of the daughter nuclei \cite{tuli}. Table \ref{table56} summarizes our results on the $\beta$ decay of $^{56}$Zn, giving the energies and intensities of the proton and $\gamma$ peaks, the $\beta$ feedings, the $B$(F) and $B$(GT) strengths (where we used $Q_{\beta}$ = 12870$^\#$(300$^\#$) keV \cite{Audi2003}, see Section \ref{masses}).

The $^{56}$Zn ground state decays by a Fermi transition to its IAS in $^{56}$Cu. From there, two $\gamma$ rays of 861 and 1835 keV are emitted, populating the 2.661 and 1.691 MeV levels in $^{56}$Cu, respectively. Due to the low $S_p$, these levels are still proton-unbound and thereafter they both decay by proton emission. Consequently the rare and exotic $\beta$-delayed $\gamma$-proton decay has been detected for the first time in the $fp$ shell \cite{PhysRevLett.112.222501}. Besides these two branches, there is a third case of a $\beta$-delayed $\gamma$-proton sequence: the 1.691 MeV level emits a $\gamma$ ray of 309 keV, going to the level at 1.391 MeV that is again proton-unbound and de-excites by proton emission.

The consequences of the $\beta$-delayed $\gamma$-proton decay on the determination of the $B$(GT) strength near the proton drip-line have been analyzed in Ref. \cite{PhysRevLett.112.222501}. These findings demonstrate that it is crucial to employ $\gamma$ detectors in such studies, stressing in particular the importance of carefully correcting the $\beta$ intensity extracted from the proton transitions for the amount of indirect feeding coming from the $\gamma$ de-excitation.

Competition between $\beta$-delayed proton emission and $\beta$-delayed $\gamma$ de-excitation is observed from the 3.508 MeV IAS in $^{56}$Cu. The $\gamma$ decays represent 56(6)\% of the total decays. The de-excitation of this $T$ = 2, $J^\pi = 0^+$ IAS via proton decay to the ground state of $^{55}$Ni ($T$ = 1/2, $J^\pi = 7/2^-$) is isospin-forbidden. Therefore the 44(6)\% proton emission that we observe can only happen because of a $T$ = 1 isospin impurity. Moreover, we have found evidence for fragmentation of the Fermi strength due to strong isospin mixing with a 0$^{+}$ state lying inside the 3.423 MeV peak \cite{PhysRevLett.112.222501}. The isospin impurity in the $^{56}$Cu IAS, $\alpha^2$ = 33(10)\%, and the off-diagonal matrix element of the charge-dependent part of the Hamiltonian, \mbox{$\left\langle{H_c}\right\rangle$ = 40(23) keV}, responsible for the isospin mixing of the 3.508 MeV IAS (mainly $T$ = 2, $J^\pi = 0^+$) and the 0$^{+}$ part of the 3423 keV level (mainly $T$ = 1), agree with the values obtained in the mirror nucleus $^{56}$Co \cite{PhysRevC.88.054329}.

\begin{table*}[!t]
	\caption{Summary of the results for the $\beta^{+}$ decay of $^{56}$Zn. Centre-of-mass proton energies $E_p$, $\gamma$-ray energies $E_{\gamma}$, and their intensities (normalized to 100 decays) $I_p$ and $I_{\gamma}$, respectively. Excitation energies $E_X$, $\beta$ feedings $I_{\beta}$, Fermi $B$(F) and Gamow-Teller $B$(GT) transition strengths to the $^{56}$Cu levels.}
	\label{table56}
	\centering
	\begin{ruledtabular}
	  \begin{tabular}{r r r r r r r r}
	 $E_p$(keV) & $I_p$(\%)& $E_{\gamma}$(keV)& $I_{\gamma}$(\%)& $E_X$(keV)& $I_{\beta}$(\%) & $B$(F)  & $B$(GT) \\	\hline
			2948(10)    & 18.8(10) &   1834.5(10)     &      16.3(49)      & 3508(140)$^a$	&	  43(5)         &    2.7(5)  &                  \\
		                      &                 &     861.2(10)     &        2.9(10)      & 			                &                     &                 &                  \\
		  2863(10)    & 21.2(10) &                           &                           & 3423(140)   &   21(1)         &    1.3(5)  & $\leq$0.32\\
		  2101(10)    & 17.1(9)   &                           &                           & 2661(140)   &   14(1)         &                & 0.34(6)      \\
		  1977(10)    &   4.6(8)   &                           &                           & 2537(140)   &   	0              &                & 0                 \\
		  1131(10)    & 23.8(11) &     309.0(10)     &                           & 1691(140)   &   22(6)         &                & 0.30(9)       \\
		    831(10)    &   3.0(4)    &                          &                           & 1391(140)    &     0              &                & 0                  \\
		\end{tabular}
	\end{ruledtabular}
	\raggedright{$^a$ Main component of the IAS}
	\vspace{-5.0 mm}
\end{table*}

Thus, the proton decay of the IAS proceeds thanks to its $T$ = 1 component. However, considering the quite large isospin mixing in $^{56}$Cu, the much faster proton decay ($t_{1/2}\sim~10^{-18}$ s) should dominate the $\gamma$ de-excitation ($t_{1/2}\sim~10^{-14}$ s in the mirror). This is not the case since we still observe the $\gamma$ decay of the IAS in competition with it. Knowledge of the nuclear structure of the three nuclei involved in the decay ($^{56}$Zn, $^{56}$Cu and $^{55}$Ni) may provide us with a possible explanation for the hindrance of the proton decay, as discussed in Ref. \cite{Rubio}.

Shell model calculations are in progress to corroborate these ideas \cite{Poves}. The preliminary results give a spectroscopic factor of 10$^{-3}$ for the proton decay of the $T$ = 1 component of the IAS to the ground state of $^{55}$Ni. They also confirm the amount of isospin mixing observed.

\section{\label{masses}Determination of the masses}

A knowledge of the masses of the proton-rich nuclei under study and their daughters is important for the determination of some key quantities. The difference between the mass excesses of the parent and $\beta$-daughter nuclei gives the $Q_{\beta}$ value of the decay, which enters in the determination of the $\beta$-decay strengths (see Section \ref{strength}). In addition, the information on the mass excess of the proton-daughter after the $\beta$ decay and of the $\beta$-daughter allows the calculation of the proton separation energy $S_p$ which, together with the measured proton energy $E_p$, provides the excitation energy of the levels populated in the daughter nucleus as $E_X=E_p + S_p$.  

If the mass excesses of at least three members of an isospin multiplet are known then the mass excess of the remaining member can be determined from the Isobaric Multiplet Mass Equation (IMME) \cite{IMMEpaper,IMME,MacCormick201461}:
\begin{equation}
  \Delta m(\alpha,T,T_z) = a + b~T_z + c~T_z^2\,,
  \label{Eq9}
\end{equation}

In the present section we determine the mass excesses of the $T_z$ = -2 nuclei $^{48}$Fe, $^{52}$Ni and $^{56}$Zn using the IMME with four members of each quintuplet.

For the $A$ = 48, $T$ = 2, $J^{\pi}$ = 0$^{+}$ mass multiplet we consider the following mass excesses: the $^{48}$Ti ground state (-48491.7(4) keV \cite{Audi2012}); the IAS in $^{48}$V (-41458.8(14) keV), obtained from the ground state mass in Ref. \cite{Audi2012} and the most recent measurement of the IAS excitation energy \cite{Ela2015}; the IAS in $^{48}$Cr (-34067(17) keV \cite{Audi2012,NDS2006}); the IAS in $^{48}$Mn (-26254(12) keV), which we determine from the mass excess of $^{47}$Cr \cite{Audi2012} and our measured \mbox{$E_p$ = 1018(10) keV} (see Table \ref{table48}). We obtain for $^{48}$Fe a mass excess of -18088(15) keV.

For the $A$ = 52, $T$ = 2, $J^{\pi}$ = 0$^{+}$ mass multiplet we consider the following mass excesses: the $^{52}$Cr ground state (-55418.1(6) keV \cite{Audi2012}); the IAS in $^{52}$Mn (-47780.9(18) keV), obtained from the ground state mass in Ref. \cite{Audi2012} and the IAS excitation energy taken from Ref. \cite{NDS2015} (combining the measurements of Refs. \cite{PhysRevC.7.677,Meyer}); the IAS in $^{52}$Fe (-39771(9) keV \cite{Audi2012,NDS2015}); the IAS in $^{52}$Co (-31561(14) keV), which we determine from the mass excess of $^{51}$Fe \cite{Audi2012} and our measured $E_p$ = 1352(10) keV (see Table \ref{table52}). We obtain for $^{52}$Ni a mass excess of -22916(16) keV.

For the $A$ = 56, $T$ = 2, $J^{\pi}$ = 0$^{+}$ mass multiplet we consider the mass excesses of the $^{56}$Fe ground state (\mbox{-60606.4(5) keV} \cite{Audi2012}) and the IAS in $^{56}$Ni (-43963(4) keV \cite{Audi2012}). We also take the mass excess of the IAS in $^{56}$Co (-52461(2) keV), obtained from the ground state mass in Ref. \cite{Audi2012} and the high-resolution CE measurement \cite{PhysRevC.88.054329}, where we have used the excitation energy of the unperturbed $T$ = 2 state taking into account the isospin mixing observed. Finally, we also use the mass excess of the IAS in $^{56}$Cu (-35128(66) keV), which we determine from the mass excess of $^{55}$Ni \cite{Audi2012} and our measured $E_p$ = 2948(10) keV (see Table \ref{table56}), where similarly we have calculated the unperturbed energy for the $T$ = 2 state considering the isospin mixing. We obtain for $^{56}$Zn a mass excess of -25911(20) keV.

\begin{figure}[!h]
  \centering
  \includegraphics[width=1\columnwidth]{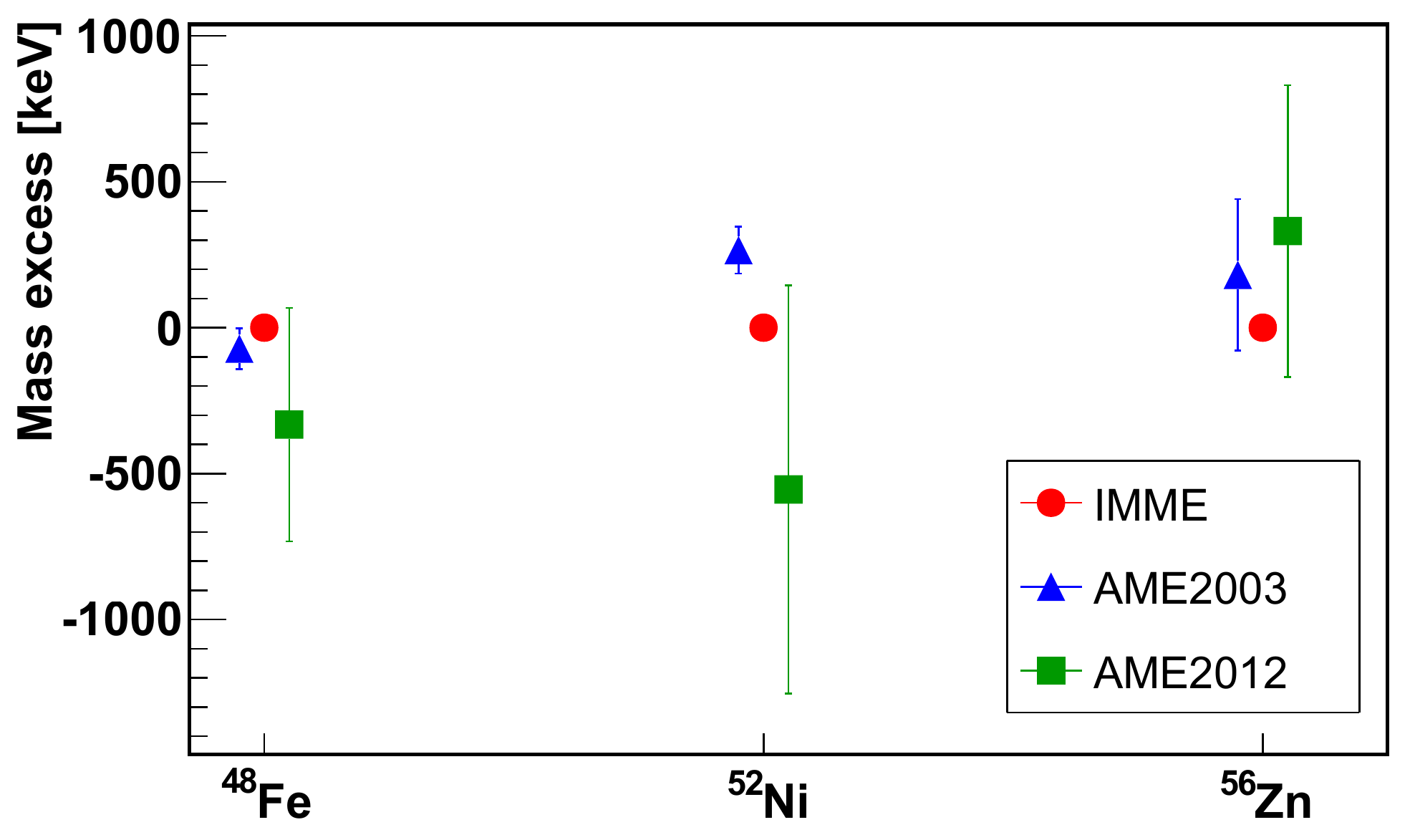}
  \vspace{-5.0 mm}
	\caption{Mass excesses of the $T_z$ = -2 nuclei $^{48}$Fe, $^{52}$Ni and $^{56}$Zn. The difference is shown between the values we obtain from the IMME (red circles) and values from the systematics 2003 AME \cite{Audi2003} (blue triangles) and 2012 AME \cite{Audi2012} (green squares). The data points belonging to each nucleus are slightly displaced to show the error bars better.}
	\label{mass}
  \vspace{-5.0 mm}
\end{figure}

The input values for the IMME calculations are summarized in Table \ref{masstab} together with the mass excesses determined above for $^{48}$Fe, $^{52}$Ni and $^{56}$Zn. In the same table these mass excesses are compared to the values from the 2003 \cite{Audi2003} and 2012 \cite{Audi2012} Atomic Mass Evaluations (AME), which are all deduced from systematics. The comparison is also shown in Fig. \ref{mass}, where we plot the differences between our IMME values and the 2003 AME and 2012 AME values. For all the nuclei, the values from the 2003 AME lie much closer to our IMME estimates than the 2012 AME ones, which agree only thanks to the larger error bars. A measurement of these masses would be important to constrain the future AME. Ref. \cite{DelSanto2014453} reports similar issues related to the 2012 AME entries for $^{66}$Se and $^{70}$Kr.

\begin{table*}[!t]
	\caption{Mass excesses (in keV) of the four members of each $T$ = 2 mass multiplet, which are used as the input for the IMME calculations. Mass excesses (in keV) of the $T_z$ = -2 nuclei $^{48}$Fe, $^{52}$Ni and $^{56}$Zn obtained by the IMME calculations and compared with the 2003 AME \cite{Audi2003} and 2012 AME \cite{Audi2012} systematics.}
	\label{masstab}
	\centering
	\begin{ruledtabular}
	  \begin{tabular}{c c c c c c c c}
	 Mass & \multicolumn{4}{c}{$T$ = 2 input values} & IMME results & 2003 AME & 2012 AME \\
	 multiplet & \multicolumn{4}{c}{ } & (this work) & \cite{Audi2003} & \cite{Audi2012} \\
	 \cline{2-5} \cline{6-6} \cline{7-8}  
	  & $T_z$ = +2 & $T_z$ = +1 & $T_z$ = 0 & $T_z$ = -1 & $T_z$ = -2 & $T_z$ = -2 & $T_z$ = -2 \\ \hline
	$A$ = 48 & -48491.7(4) \cite{Audi2012} & -41458.8(14) & -34067(17) \cite{Audi2012,NDS2006} & -26254(12) & -18088(15) & -18160$^\#$(70$^\#$) & -18420$^\#$(400$^\#$) \\
	$A$ = 52 & -55418.1(6) \cite{Audi2012} & -47780.9(18) & -39771(9) \cite{Audi2012,NDS2015} & -31561(14) & -22916(16) & -22650$^\#$(80$^\#$) & -23470$^\#$(700$^\#$) \\
	$A$ = 56 & -60606.4(5) \cite{Audi2012} & -52461(2)$^a$ & -43963(4) \cite{Audi2012} & -35128(66)$^a$ & -25911(20) & -25730$^\#$(260$^\#$) & -25580$^\#$(500$^\#$) \\
		\end{tabular}
	\end{ruledtabular}
	\raggedright{$^a$ Based on the unperturbed energy of the $T$ = 2 state. $^\#$ Values obtained from systematics.}
  \vspace{-2.0 mm}
\end{table*}

Among the three nuclei under investigation, $^{48}$Fe is the only case where the ground state mass of the daughter $^{48}$Mn can be determined directly from the measurements, hence the $Q_{\beta}$ value and $S_p$ can be determined without further assumptions. This is not possible for $^{52}$Ni and $^{56}$Zn, where no mass measurement for the daughter exists and therefore, in view of the rather large discrepancies observed in Fig. \ref{mass}, we have to rely on mirror symmetry.

$^{48}$Fe. Starting from the mass of the IAS in $^{48}$Mn (-26254(12) keV, see above), the ground state mass of $^{48}$Mn can be determined by subtracting the energies of the de-exciting $\gamma$-ray cascade (see Table \ref{table48}). We obtain -29290(12) keV, which agrees with the 2012 AME measured value of -29320(170) keV. Finally, using our derived (more precise) value, we calculate $Q_{\beta}$ = 11202(19) keV for the decay of $^{48}$Fe and $S_p$ = 2018(10) keV in $^{48}$Mn.

$^{52}$Ni. In order to determine the ground state mass of $^{52}$Co, we can again start from the IAS in $^{52}$Co (\mbox{-31561(14) keV}, see above) and subtract the energies of the observed de-exciting $\gamma$ rays (see Table \ref{table52}) up to the first excited 2$^+$ state in $^{52}$Co. At this point we have assumed an energy of 378 keV for this 2$^+$ state, which we take from the value in the mirror nucleus $^{52}$Mn, 377.749(5) \cite{NDS2015,PhysRevC.7.677}, where we estimated an error of 50 keV by looking at the energies of the levels up to 400 keV in mirror nuclei with $T_z$ = +1/2, -1/2, +1, -1. In this way we obtain a ground state mass excess of \mbox{-34487(52) keV} in $^{52}$Co. For comparison, the mass excess for $^{52}$Co is \mbox{-33990$^\#$(200$^\#$) keV} in the 2012 AME \cite{Audi2012} and \mbox{-33920$^\#$(70$^\#$) keV} in the 2003 AME \cite{Audi2003}, where $^\#$ indicates values derived from systematics. Based on our deduced mass excesses for $^{52}$Co and $^{52}$Ni, we calculate \mbox{$Q_{\beta}$ = 11571(54) keV} for the decay of $^{52}$Ni and \mbox{$S_p$ = 1574(51) keV} in $^{52}$Co.

$^{56}$Zn. For $^{56}$Zn we presented in Ref. \cite{PhysRevLett.112.222501} the complete strength calculations for both AMEs. The difference in the $B$(F) and $B$(GT) strengths obtained with the two AMEs is tiny (see Table II of Ref. \cite{PhysRevLett.112.222501}) and the values agree well within the uncertainties. However, we already noticed that the energies of the mirror levels in $^{56}$Cu and $^{56}$Co agree within 100 keV when the 2003 AME \cite{Audi2003} is used, while they differ by $\sim400$ keV if one uses the 2012 AME \cite{Audi2012}. Therefore we argued that the 2003 AME gives a more reasonable value for the energy of the IAS. The IMME calculation for the mass excess of the $^{56}$Zn ground state confirms that the systematic value from the 2003 AME lies closer to the $expected~value$ than the 2012 AME one. Hence in the present paper we continue to prefer the 2003 AME \cite{Audi2003} values and so we use \mbox{$Q_{\beta}$ = 12870$^\#$(300$^\#$) keV} for the decay of $^{56}$Zn and \mbox{$S_p$ = 560$^\#$(140$^\#$) keV} in $^{56}$Cu.

\section{\label{concl}Conclusions}

In the present paper we reported on a study of the $\beta$ decays of the $^{48}$Fe, $^{52}$Ni and $^{56}$Zn nuclei, performed at GANIL. These three exotic nuclei lie close to the proton drip-line and have a third component of isospin $T_z$ = -2.

In the first half of the paper we have extensively described the experiment and data analysis procedures, taking $^{56}$Zn as an example and discussing differences in the analysis of $^{48}$Fe and $^{52}$Ni. The results obtained for each of the three nuclei have been presented in the second half of the paper.

We have extracted the half-lives and the total $\beta$-delayed proton emission branching ratios for all of the nuclei under study. Individual $\beta$-delayed protons and $\beta$-delayed $\gamma$ rays have been measured and the related branching ratios have been determined, most of them for the first time. Partial decay schemes have been determined for the three nuclei. The higher energy resolution achieved for protons, in comparison to previous studies, allowed us to identify new states populated in the decays of $^{48}$Fe and $^{52}$Ni and to establish for the first time the partial decay scheme of $^{56}$Zn. We have used the Isobaric Multiplet Mass Equation to deduce the mass excesses of the nuclei under study using information from our experimental data and the literature. Moreover we have determined the absolute Fermi and Gamow-Teller transition strengths. 

In $^{48}$Fe, $^{52}$Ni and $^{56}$Zn the de-excitation of the $T$ = 2 IAS via $\beta$-delayed proton emission is isospin-forbidden, however it is observed in all cases. This is attributed to a $T$ = 1 isospin impurity in the IAS wave function. Furthermore, we have observed in all three nuclei competition between $\beta$-delayed protons and $\beta$-delayed $\gamma$ rays. Nevertheless, while for $^{48}$Fe and $^{52}$Ni the $\beta$-delayed $\gamma$ de-excitation of the IAS dominates (86(19) \% and 75(23) \%, respectively), in $^{56}$Zn the $\gamma$ decays are only 56(6) \% of the total decays from the IAS.

The case of $^{56}$Zn is, indeed, peculiar for various reasons. The comparison with the mirror CE experiment shows, only for this nucleus, that there is another 0$^{+}$, $T$ = 1 state which mixes at 33\% with the IAS, making the proton decay allowed. Another interesting feature is that in $^{48}$Fe and $^{52}$Ni the calculated partial proton half-lives are of the same order-of-magnitude as the $\gamma$-decay Weisskopf transition probabilities (this is partly due to the relatively low proton energies involved). In contrast, in $^{56}$Zn the proton decay is expected to be four orders-of-magnitude faster than the $\gamma$ de-excitation. However, the latter is still observed. This indicates some hindrance of the proton decay. The explanation may lie in nuclear structure reasons, as explained in Ref. \cite{Rubio}. Shell model calculations are in progress \cite{Poves} and the preliminary results confirm the amount of isospin mixing experimentally observed and the hindrance of the proton decay.

\begin{acknowledgments}
This work was supported by the Spanish MICINN grants FPA2008-06419-C02-01, FPA2011-24553; Centro de Excelencia Severo Ochoa del IFIC SEV-2014-0398; CPAN Consolider-Ingenio 2010 Programme CSD2007-00042; $Junta~para~la~Ampliaci\acute{o}n~de~Estudios$ Programme (CSIC JAE-Doc contract) co-financed by FSE; ENSAR project 262010; MEXT, Japan 18540270 and 22540310; Japan-Spain coll. program of JSPS and CSIC; UK Science and Technology Facilities Council (STFC) Grant No. ST/F012012/1; Region of Aquitaine. E.G. acknowledges support by TUBITAK 2219 International Post Doctoral Research Fellowship Programme. R.B.C. acknowledges support by the Alexander von Humboldt foundation and the Max-Planck-Partner Group. We acknowledge the EXOGAM collaboration for the use of their clover detectors. We thank Prof. Ikuko Hamamoto and Dr. J. L. Ta{\'i}n for useful discussions.
\end{acknowledgments}

\bibliography{references}

\end{document}